\documentclass{article}
\usepackage[dvipsnames]{xcolor}
\usepackage{authblk}
\usepackage{float}
\usepackage{bm}
\usepackage{placeins}
\usepackage{arxiv}
\makeatletter
\renewcommand{\@date}{}         % removes date under the title
\makeatother
\usepackage{adjustbox}
\usepackage{etoolbox}
\usepackage{amsmath}
\usepackage{bbm}
\usepackage{amsthm}
\newtheorem{definition}{Definition}
\newtheorem{remark}{Remark}
\usepackage[utf8]{inputenc} % allow utf-8 input
\usepackage[T1]{fontenc}    % use 8-bit T1 fonts
\usepackage{hyperref}       % hyperlinks
\usepackage{url}            % simple URL typesetting
\usepackage{booktabs}       % professional-quality tables
\usepackage{amsfonts}       % blackboard math symbols
\usepackage{nicefrac}       % compact symbols for 1/2, etc.
\usepackage{microtype}      % microtypography
\usepackage{lineno}
\usepackage{lipsum}
\usepackage{subcaption}
\usepackage{verbatim}
\usepackage{amsmath}
\usepackage{amsthm}
\usepackage{mathptmx} 
\usepackage{mathtools}
\usepackage[english]{babel}
\usepackage{hyperref}
\usepackage{mathptmx}

\numberwithin{equation}{section}
\usepackage{amssymb}
\usepackage[ruled, lined, linesnumbered, commentsnumbered, longend]{algorithm2e}
\usepackage{amsthm}
\usepackage{graphicx}
\usepackage{dsfont}
\usepackage{epstopdf} 
\usepackage[utf8]{inputenc}
\usepackage[english]{babel}
\usepackage{hyperref}
\DeclareMathAlphabet{\mathpzc}{OT1}{pzc}{m}{it}
\usepackage{mathrsfs}
\usepackage{amsbsy}
\usepackage{amssymb} 
\usepackage{algpseudocode}
\usepackage{listings}
\MakeRobust{\Call}
\allowdisplaybreaks
\usepackage{authblk}  % for affiliations

\title{
A Physics-Augmented Machine Learning Constitutive Model for Damage in Solids}

\author[1]{\Large Amirhossein Amiri-Hezaveh}
\author[2]{\Large Adrian Buganza Tepole}

\affil[1]{\small \textit{Department of Mechanical Engineering, Purdue University, West Lafayette, IN, USA}}
\affil[2]{\small \textit{Department of Mechanical Engineering, Columbia University, NY 10027, USA}}

\begin{document}
\date{{}}
\maketitle
\footnotetext[2]{ \texttt{a.buganzatepole@columbia.edu}}
\begin{abstract}
We propose a data-driven constitutive framework for anisotropic damage mechanics based on the second-order damage tensor approach for both compressible and incompressible materials. The formulation is thermodynamically consistent in the sense of satisfying the Clausius-Duhem inequality. The strain energy density potentials are expressed as isotropic functions of the right Cauchy–Green deformation tensor, along with structural tensors that encode anisotropy that is either present in the virgin material or resulting from damage. To guarantee the polyconvexity condition, non-decreasing convex neural networks with inputs that ensure polyconvexity are used to parameterize the strain energy density potentials. The model vanishes in the undeformed state fulfilling the normality condition. In contrast to classical [1-d] damage models, the expressiveness of the new data-driven model is enhanced by employing a family of nonlinear, convex, decreasing functions to capture the effect of damage. The damage evolution is governed through damage potential where the corresponding damage threshold is in terms of the damage conjugate forces. Furthermore, as a special case of the general formulation, a new anisotropic generic format is introduced to predict constitutive responses under damage-induced anisotropy in initially isotropic materials. To address the computational burden associated with training due to the dissipative nature of the problem, a new decoupled training scheme is introduced, whose accuracy is tested and demonstrated in all numerical examples.       Numerical examples, including synthetic benchmarks for incompressible isotropic and transversely isotropic materials as well as compressible orthotropic material, are provided to evaluate the performance of the method. Finally, the framework is validated against experimental data capturing anisotropic Mullins-type damage. As indicated by numerical results, the new data-driven model effectively captures the damage behavior of materials while adhering to the physical and mathematical constraints of constitutive modeling.                     
\end{abstract}
% keywords can be removed
%\keywords{..... }
\section{Introduction}
Modeling the progressive degradation of soft materials under large deformations remains a significant challenge in computational mechanics \cite{tac2024data}. Traditional constitutive models often rely on phenomenological assumptions or micromechanical idealizations that struggle to capture the anisotropic and path-dependent nature of damage observed in biological tissues, polymers, and elastomers \cite{ricker2021comparison}. In particular, the emergence of anisotropy induced by damage, such as in the Mullins effect, is difficult to describe using classical frameworks based on predefined energy forms and limited internal variables. There is thus a pressing need for constitutive models that are both flexible enough to represent complex damage mechanisms and grounded in thermodynamic consistency. Recent advances in data-driven modeling offer a promising avenue to address this gap by combining physics-based priors with machine learning architectures that can adapt to experimental data with minimal assumptions \cite{fuhg2024review}. In this work, we develop a thermodynamically consistent, data-driven model for anisotropic damage using a second-order damage tensor and convex neural networks, with particular emphasis on capturing damage-induced anisotropy in soft materials.

\textit{Damage} is a form of irreversible behavior observed in materials, defined as the degradation of mechanical resistance due to the development of defects. Continuum Damage Mechanics (CDM) is the continuum-based approach used to quantify such degradations. The earliest attempts to model damage date back to \cite{kachanov1999rupture, Rabotnov1969,hayhurst1973effect}, where damage was modeled as the reduction of the effective area as a result of nucleation and coalescence of micro-cracks/cavities through a scalar variable. However, the degradation of material properties is not only limited to micro-cracks. For example, the Mullins' effect \cite{holt1932behavior,mullins1948effect,mullins1957theoretical}, an observed stress softening during loading and unloading cycles that was initially seen in rubbers and then in other materials such as double network hydrogels \cite{gong2003double,mai2017induced,mai2018distinctive} and multiple network elastomers \cite{ducrot2014toughening,millereau2018mechanics} are another type of damage. In fact, the Mullins' effect in polymeric materials is the result of the loss of bonds between polymer chains and the matrix or the rupture of polymer chains (see, for example, \cite{govindjee1991micro}). \\

The scalar representation of the damage variable results in isotropic stress softening, meaning a uniform degradation of mechanical properties in all directions. Some of the key contributions to scalar damage modeling include \cite{leckie1978constitutive}, \cite{cordebois1982anisotropic}, \cite{cordebois1982damage}, \cite{lemaitre1984use}, \cite{Lemaitre1985Continuous}, \cite{mazars1985application}, \cite{simo1987strain}, \cite{ju1989energy}, and \cite{carol1994unified}.  For instance, the concept of equivalent (complementary) energy, in which the net stress in the fictitious undamaged material is defined such that the energy of the fictitious undamaged material equals that of the damaged one, was introduced in \cite{cordebois1982anisotropic,cordebois1982damage}. Additionally, the formulation of continuum damage models based on strain (involving the concepts of effective stress and equivalent strain) and stress (involving effective strain and equivalent stress) was presented in \cite{simo1987strain}. In this work, motivated by the definition of the conjugate thermodynamic force associated with the damage variable, the damage state is characterized by a potential function where the damage threshold is in terms of elastic stored energy, leading to symmetric elastic-damage moduli.\\

However, in general, the damage is \textit{directional}. As justified in \cite{leckie1981tensorial} and later in \cite{onat1988representation}, to account for directional degradations, the so-called anisotropic damage, the corresponding internal variable should contain information about the magnitude and direction simultaneously, which can be represented via the even-rank tensorial form. In this regard, one way to account for anisotropic damage is through the degradation of the stiffness tensor, which can be interpreted as a special case of a linear transformation by means of a fourth-order tensor that maps the homogenized stress tensor to the effective stress tensor (\cite{sidoroff1981description}, \cite{cordebois1982anisotropic}, \cite{Ladeveze1983},\cite{simo1987strain},\cite{chow1987anisotropic},\cite{ju1989energy}, \cite{ju1990isotropic}, \cite{maire1997new}, \cite{fichant1997continuum}, \cite{carol2001formulation}). Hence, these models are good candidates when the change in tangent stiffness or compliance tensors is governed only by the state of damage and not because of material or geometrical nonlinearities. \\

Among tensorial forms, the second-order damage tensor has received attention due to its rather tangible physical interpretation and the capability of modeling orthotropic damage, which can be considered a satisfactory anisotropic model (see, for example, \cite{murakami2012continuum} (pp. 26) and references therein). An extension of the concept of scalar damage to second-order damage tensor, interpreted in terms of area reduction due to micro-crack growth and nucleation, can be found in \cite{murakami1981continuum, Murakami1988}. In these works, the effect of damage is represented through the concept of a fictitious undamaged continuum body along with tensorial internal variables. The resulting effective stress is asymmetric, and thus,  only its symmetric part is considered. This symmetrization can be interpreted as a special case of the definition of the fourth-order damage tensor \cite{cordebois1982damage} that maps a symmetric second-order tensor to a net symmetric stress tensor. \\

To model nonlinear constitutive equations for isotropic hyperelastic materials, the Helmholtz free energy potential can be expressed in terms of the invariants of the right Cauchy-Green tensor. However, representing anisotropy in nonlinear hyperelastic constitutive equations is nontrivial. In this regard, the elegant idea of expressing an anisotropic function in terms of isotropic functions was introduced in \cite{boehler1979simple} and later developed in \cite{liu1982representations} (see also \cite{10.1115/1.3111066} for further details). This concept was further advanced in \cite{zhang1990structural} (see also \cite{itskov2009tensor}), where it was shown that any scalar anisotropic function, in which a set of structural tensors characterizes anisotropy, can be represented by an isotropic function with the structural tensors included as additional arguments. This idea has been successfully used in several works, including, for example,  \cite{schroder2003invariant} \cite{itskov2004class,schroder2008anisotropic}.  In parallel, as noted in \cite{10.1115/1.3111066} and applied in \cite{matzenmiller1994damage}, anisotropy induced by an irreversible process, such as damage, can be modeled in a similar manner by considering the corresponding internal variables as structural tensors. In this vein, a geometric method to account for damage in the nonlinear regime was developed in   \cite{steinmann1998framework}, which is based on the assumption of fictitious undamaged material developed in \cite{Murakami1988} and the energy equivalence principle \cite{cordebois1982damage}.  Subsequently, the computational aspect of the method was rigorously set up in \cite{menzel2001theoretical}. Moreover, recently, motivated by the aforementioned works and \cite{svendsen2001modelling}, a new nonlinear damage combined with plasticity has been proposed in \cite{reese2021using}. \\

The aforementioned studies have primarily relied on the closed-form representation of the Helmholtz potential and internal variables. However, data-driven methods have demonstrated significant flexibility in predicting material behavior, owing to the universal approximation property of neural networks \cite{hornik1989multilayer} and the emergence of advanced machine learning frameworks such as PyTorch, TensorFlow \cite{tensorflow2015-whitepaper}, and JAX \cite{jax2018github}. In this work, we focus on damage modeling, while referring the reader to \cite{fuhg2024review} for a comprehensive discussion on data-driven constitutive models in solid mechanics. Among data-driven methods, physics-based data-driven methods have received the primary attention as the results are consistent with the laws of physics either by construction or through a weak sense as a penalty term in the loss functions. Some of the major contributions to the physics-based constitutive equations can be found in:\cite{vlassis2020geometric,linka2021constitutive,chen2022polyconvex,as2022mechanics,fuhg2022learning,kalina2023fe,tac2022data,linden2023neural,linka2023new,tacc2024benchmarking}. Among the contributions above, the invariants-based constitutive relations that are used as input arguments for either the convex of function input neural networks (ICNN) \cite{amos2017input}and neural ordinary differential equations (NODEs) \cite{chen2018neural} to satisfy polyconvexity condition (\cite{ball1976convexity, schroder2003invariant}, a mathematical condition that together with coercivity condition ensure the ellipticity conditions (a physical condition). In the context of data-driven constitutive equations incorporating damage, Mullin's effect is considered through an isotropic damage variable in \cite{zlatic2024recovering}. However, as the Mullins' effect in general induces anisotropy \cite{mullins1948effect, mai2017induced,mai2018distinctive}, it would be more desirable to develop a model that incorporates induced anisotropy.\\

In this paper, a new data-driven method for measuring anisotropic damage in materials is proposed. The method can be considered a generalization of isotropic damage introduced in \cite{simo1987fully} and our previous isotropic data-driven framework \cite{tac2024data}. To account for anisotropic damage, we consider internal damage variables as  second-order tensors. As explained in the sequel, several possibilities exist to formulate the damage problem depending on the initial degree of anisotropy of virgin material and the anisotropy induced by damage. Hence, for simplicity, we formulate the problem based on the following assumption: as the degree of anisotropy of the material, we assume that the virgin material is at most \textit{orthotropic}, where we additionally assume the axes of material symmetry align with the principal axes of the damage tensor.  The resulting formulation can be straightforwardly generalized for when the material symmetry is not necessarily aligned with the principal directions of the damage by adding structural tensors used to define the initial anisotropy and those of the damage principal axes (see, for example,  \cite{reese2021using}). The formulation is invariant-based, where, following the approach in \cite{itskov2004class}, anisotropy is incorporated in accordance with the structural tensor approach stated in \cite{zhang1990structural}.To guarantee polyconvexity, we utilize invariants that are convex functions of the standard polyconvex arguments—namely, the deformation gradient, its cofactor, and the Jacobian determinant—as input features to a generic non-decreasing ICNN, which is parameterized during training. This construction fulfills objectivity, anisotropy, and polyconvexity identically.  By defining the stress and the thermodynamic forces associated with damage as derivatives of the Helmholtz free energy with respect to their conjugate variables, and by employing an evolution procedure analogous to that developed in \cite{simo1987fully} for damage state equations expressed in terms of thermodynamic conjugate forces, the formulation ensures thermodynamic consistency. It is noted that new method is no longer restricted to classical [1-d] format as it incorporates attenuation functions into the Helmholtz free energy potential via a weighted sum of decreasing, convex functions, with both the weights and cutoff coefficients treated as trainable functions, as discussed in the sequel.To ensure the normality condition for the Helmholtz free energy and its associated stress response, we adopt the approach of \cite{linden2023neural}, incorporating additional consistent terms into the energy function. As a special case of the general formulation, we present a generic model applicable to virgin isotropic materials, where anisotropy arises through the evolution of damage. In addition, we introduce a novel and scalable framework that significantly reduces the computational cost of training. The formulation's validity is demonstrated through four examples: three synthetic datasets and an experimental dataset from \cite{mai2018distinctive}, which characterizes the damage-induced anisotropy in double network hydrogels. For synthetic data, we consider two incompressible examples, one isotropic and one transversely isotropic cases. As third synthetic example, we consider a general format  for discovering stress-strain curves corresponding to a synthetic compressible material.         

% \cite{simo1987fully,itskov2004class, Murakami1988,zhang1990structural,svendsen2001modelling,reese2021using}. In particular, the Helmholtz free energy is 
\section{Background}\label{back}
Several considerations are required to develop a constitutive equation incorporating damage in materials through internal state variables \cite{coleman1967thermodynamics}.To begin with, we assume that a continuum deformable body, denoted by $\mathcal{B}^0$, is deformed to the spatial configuration $\mathcal{B}^t$. In this regard, the deformation gradient and the right Green-Cauchy strain are defined as ${\bf{F}} = {\nabla _{\bf{X}}}{\bf{x}}$ and ${\bf{C}} = {{\bf{F}}^T}{\bf{F}}$, respectively, with  ${\bf{x}}$ and ${\bf{X}}$ are spatial and material coordinates. We now briefly discuss the physical and mathematical conditions that are employed for the formulation of the new data-driven damage constitutive equations.
\subsection{Objectivity}
Objectivity is a physical condition that makes the resulting constitutive equation frame-indifferent. According to this condition,  any constitutive equations must be invariant under any rigid body motions. For hyper-elastic materials, this requirement leads to:
\begin{linenomath*}
\begin{equation}\label{b0}
\begin{aligned}
{{\bar \psi }^e}({\bf{QF}}) = {{\bar \psi }^e}({\bf{F}})
\end{aligned}
\end{equation}
\end{linenomath*}
where ${{\bar \psi }^e}$ stands for the Helmholtz free energy. The implication of the above condition is  ${{\bar \psi }^e} = {{\bar \psi }^e}({\bf{C}})$. Namely, the Helmholtz free energy can be assumed to depend on the objective strain metric $\mathbf{C}$.
\subsection{Thermodynamics}
  The laws of thermodynamics together have direct implications in the derivation of constitutive equations, which we briefly discuss here. The first law of thermodynamics states the conservation of energy, which has the following form:
\begin{linenomath*}
\begin{equation}\label{b1}
\begin{aligned}
\int\limits_{{\mathcal{B}^t}} {\rho \dot e d\Omega }  = \int\limits_{{\mathcal{B}^t}} {\left( {{\mathbf{\sigma }}:{\text{sym}}\left( {{\nabla _{\mathbf{x}}}{\mathbf{\dot u}}} \right) - \nabla  \cdot {\mathbf{q}} + r} \right) d\Omega }, 
\end{aligned}
\end{equation}
\end{linenomath*}
where ${\text{sym}}\left( {{\nabla _{\mathbf{x}}}{\mathbf{\dot u}}} \right) = \frac{1}{2}\left( {{\nabla _{\mathbf{x}}}{\mathbf{\dot u}} + {\nabla _{\mathbf{x}}}{{{\mathbf{\dot u}}}^{\text{T}}}} \right)$, $e$, $\mathbf{q}$, and $r$ are, respectively, symmetric part for the gradient of velocity field, internal energy,  heat flux,  and heat source in the system (material), respectively. On the other hand, the second law of thermodynamics deals with the direction of the flow of energy and ensures that the  net production of entropy of an isolated system is always non-negative, the so-called Clausius–Duhem inequality:
\begin{linenomath*}
\begin{equation}\label{b2}
\begin{aligned}
H = \int\limits_{{{\mathcal{B}}^t}} {\rho \dot \eta \,d\Omega }  - \int\limits_{{{\mathcal{B}}^t}} {\frac{r}{\theta }d\Omega }  + \int\limits_{\partial {{\mathcal{B}}^t}} {{\mathbf{n}} \cdot \left( {\frac{{\mathbf{q}}}{\theta }} \right)d\Gamma }  \ge 0,
\end{aligned}
\end{equation}
\end{linenomath*}
in which $\rho \dot \eta $ is the rate of change of total entropy, and the second and third terms are, respectively,  volume and surface, rate of entropy change due to heat transfer to the system. Defining the Helmholtz energy function as ${\bar \psi ^e} = e - \eta \theta$, and considering \eqref{b1}, after some classical manipulations, we obtain the following localized form for an isothermal process in the reference coordinate system:
\begin{linenomath*}
\begin{equation}\label{b3}
\begin{aligned}
\frac{1}{2} {\bf S} \cdot {\bf \dot C} - \mathop {{{\bar \psi }^e}}\limits^.  \ge 0,
\end{aligned}
\end{equation}
\end{linenomath*} 
where $\mathbf{S}$ is the second Piola–Kirchhoff stress tensor. For an isothermal process the right Cauchy-Green tensor and a list of internal variables, denoted by $\left\{ {{{\bm{\xi}}_1},...,{{\bm{\xi}}_n}} \right\}$ are arguments of the Helmholtz free energy function, that is,  ${\bar \psi ^e} = {\bar \psi ^e}({\mathbf{C}},{{\bm{\xi}}_1},...,{{\bm{\xi}}_n})$. Hence,  one can write:
\begin{linenomath*}
\begin{equation}\label{b4}
\begin{aligned}
\left( {\frac{1}{2} \mathbf{S}  - \frac{{\partial {{\bar \psi }^e}}}{{\partial {\mathbf{C}}}}} \right){\mathbf{\dot C}} - \sum\limits_{i = 1}^n {\frac{{\partial {{\bar \psi }^e}}}{{\partial {{\bm{\xi}}_i}}}{{{\bm{\dot \xi }}}_i}}   \ge 0,
\end{aligned}
\end{equation}
\end{linenomath*}
which must be satisfied for any thermodynamic processes, implying further:
\begin{linenomath*}
\begin{equation}\label{b5}
\begin{aligned}
&{\bf{S}} = 2\frac{{\partial {{\bar \psi }^e}}}{{\partial {\bf{C}}}},\,\, {\bf{S}}^T={\bf{S}},\\
&\sum\limits_{i = 1}^n {{{\mathbf{A}}_i}{{{\mathbf{\dot {\bm{\xi}} }}}_i}}  \geqslant 0,
\end{aligned}
\end{equation}
\end{linenomath*}

where ${\bf{A}}_i$'s are conjugate thermodynamical forces corresponding to ${\bm{\xi}}_i$'s, $i=1,..., n$, defined as:
\begin{linenomath*}
\begin{equation}\label{b6}
\begin{aligned}
{{\bf{A}}_i} =  - \frac{{\partial {{\bar \psi }^e}}}{{\partial {{\bm{\xi}}_i}}}.
\end{aligned}
\end{equation}
\end{linenomath*}
Hence, we define the following:
\begin{definition}\label{th_consistency}
A constitutive model is thermodynamically consistent if \eqref{b5} and \eqref{b6} hold. 
\end{definition}
\subsection{Material symmetry}
Material symmetry is a physical property observed in the behavior of materials. That is, we say the material exhibits symmetry with respect to a group of transformations when the constitutive equations remain form-invariant under the action of such transformations. We explain the concept of anisotropy for a hyper-elastic material as the existence of the internal variables may further result in anisotropy, as shall be seen in the sequel. For a hyper-elastic material, in an isothermal process, the Helmholtz free energy function is ${\bar \psi }^e={\bar \psi }^e({\bf{C}})$.  
\begin{definition}
Given a group of symmetry, denoted by $\mathcal{G}$, we call a hyper-elastic material with  ${\bar \psi }^e={\bar \psi }^e({\bf{C}})$ is symmetric under the action of the elements of $\mathcal{G}$  when:
\begin{linenomath*}
\begin{equation}\label{b7}
\begin{aligned}
{{\bar \psi }^e}({{\bf{Q}}}{\bf{C}}{{\bf{Q}}^T}) = {{\bar \psi }^e}({\bf{C}})
\end{aligned}
\end{equation}
\end{linenomath*}
for every ${\bf{Q}} \in \mathcal{G}$. 
\end{definition}
When $\mathcal{G}$ is $\rm{O}(3)$, the material and the corresponding ${\bar \psi }^e$ are called \textit{isotropic material} and \textit{isotropic function}, respectively. As mentioned earlier, it is possible to represent an anisotropic scalar function in terms of an isotropic function whose arguments are augmented by the corresponding \textit{structural tensors} (see \cite{boehler1979simple,liu1982representations,zhang1990structural,10.1115/1.3111066}). In this regard, we follow \cite{zhang1990structural}:
\begin{definition}
A tensor (series) $\mathbb{ \tilde S}$ of arbitrary order is called a structural tensor associated with a symmetry group  $\mathcal{G}$, if $\mathcal{G}$ is the group under which $\mathbb{\tilde S}$ remains invariant, that is:
\begin{linenomath*}
\begin{equation}\label{b8}
\begin{aligned}
{\mathbf{Q}}*{\mathbb{\tilde S}} = {\mathbb{\tilde S}},\,\,\,\forall\, {\mathbf{Q}} \in \mathcal{G}
\end{aligned}
\end{equation}
\end{linenomath*}
where $*$ denotes the action of $\bf{Q}$ on $\mathbb{\tilde S}$.  
\end{definition}
For a crystalline elastic solid,  there exist  32 crystal classes and the transversely isotropic case (see \cite{schroder2008anisotropic}), which can be characterized by structural tensors.  In particular, triclinic, monoclinic, and orthorhombic symmetries can be represented completely by first and second-order structural tensors, while for tetragonal, rhombohedral, hexagonal, and cubic structural tensors of up to order six are necessary (see \cite{zhang1990structural}). It is known that any anisotropic tensorial function can be characterized by an isotropic function where structural tensors are added as the arguments of the function \cite{zhang1990structural}, which implies the following for  hyper-elastic materials: 
\begin{linenomath*}
\begin{equation}\label{b9}
\begin{aligned}
{{\bar \psi }^{e(aniso)}}({\bf{C}}) = {{\bar \psi }^{e(iso)}}({\bf{C}},{{\mathbb{\tilde S}}_1},...,{{\mathbb{\tilde S}}_n})
\end{aligned}
\end{equation}
\end{linenomath*}
where $\mathbb{S}=\left\{ {{{\mathbb{\tilde S}}_1},...,{{\mathbb{\tilde S}}_n}} \right\}$ characterize the anisotropy group $\mathcal{G}$ in the sense: 
\begin{linenomath*}
\begin{equation}\label{b10-0}
\begin{aligned}
\mathcal{G} \equiv \left\{ {{\bf{Q}} \in {\rm{O}}(3),\,{\bf{Q}} * {{\mathbb{\tilde S}}_i} = {{\mathbb{\tilde S}}_i},\,\,i = 1,...,n} \right\}.
\end{aligned}
\end{equation}
\end{linenomath*}
For an orthotropic material, given that three mutually perpendicular symmetry axes are indicated by   ${\bf{n}}_i, \,i=1,2,3$, the structural tensors are  denoted similarly to \cite{itskov2004class} as: 
\begin{linenomath*}
\begin{equation}\label{b10}
\begin{aligned}
{{\bf{L}}_i}= {\bf{n}}_i \otimes {\bf{n}}_i, \,\,\,\, i=1,2,3,
\end{aligned}
\end{equation}
\end{linenomath*}
and for transversely isotropic symmetry, given the axis of symmetry aligns with  ${\bf{n}}_1$, the above format is reduced into:
\begin{linenomath*}
\begin{equation}\label{b11}
\begin{aligned}
{{\bf{L}}_1} = {{\bf{n}}_1} \otimes {{\bf{n}}_1},\,{{\bf{L}}_2} = {{\bf{L}}_3} = \frac{1}{2} ({\bf I} - {\bf n}_1 \otimes {\bf n}_1)
\end{aligned}
\end{equation}
\end{linenomath*}
where $\bf{I}$ is the second order identity tensor.  It is noted that the following properties are implied from the above definition:
\begin{linenomath*}
\begin{equation}\label{b12}
\begin{aligned}
\sum\limits_{i = 1}^3 {{{\bf{L}}_i}}  = {\bf{I}},{\rm{tr}}({{\bf{L}}_i}) = 1,\,\,{{\bf{L}}_i}{{\bf{L}}_j} = {\bf{0}}.
\end{aligned}
\end{equation}
\end{linenomath*}
\begin{remark}
The existence of the internal variables may further induce anisotropy. To account for such anisotropy, one can follow the approach in  \cite{10.1115/1.3111066, menzel2001theoretical,svendsen2001modelling, reese2021using}, in which the internal variables are considered as additional structural tensors that define induced anisotropy. Hence, the above discussion can be extended to the case when there exists internal variables, i.e., ${\bar \psi }^e={{\bar \psi }^e}({\bf{C}},{{\bm{\xi}}_1},...,{{\bm{\xi}}_m})$. In particular, \eqref{b9} can be written for this case as:
\begin{linenomath*}
\begin{equation}\label{b13}
\begin{aligned}
{{\bar \psi }^{e\,(aniso)}}({\bf{C}},{{\bm{\xi}}_1},...,{{\bm{\xi}}_n})={{\bar \psi }^{e\,(iso)}}({\bf{C}},{{\bm{\xi}}_1},...,{{\bm{\xi}}_m},{{\mathbb{\tilde S}}_1},...,{{\mathbb{\tilde S}}_n}).
\end{aligned}
\end{equation}
\end{linenomath*}  
In what follows, for simplicity, we drop '\textit{(iso)}' for isotropic scalar function ${{\bar \psi }^{e\,(iso)}}$ and show it with the notation ${{\bar \psi }^{e}}$.  
\end{remark}
\subsection{Polyconvexity}
 The existence of the minimizer of total potential energy in nonlinear elasticity is secured when the strain energy function is quasi-convexity  \cite{Morrey1952}, which is an integral inequality condition, and satisfies coercivity condition. Polyconvexity is a mathematical notion, originally introduced in the seminal work \cite{ball1976convexity}, that ensures the quasi-convexity. This condition is stronger condition than quasi-convexity,  but it is more straightforward to be verified. Hence, in this work similar to several studies in the literature we impose this condition to the new data-driven framework.  In this regard, we follow \cite{schroder2003invariant}:
\begin{definition}
A scalar function ${\bar \psi }^e\left( \bf{F} \right)$ is polyconvex if and only if the function has the following representation:
\begin{linenomath*}
\begin{equation}\label{b14}
\begin{aligned}
{{\bar \psi }^e}= {{\bar \psi }^e}({\bf{F}},{\rm{cof }}\,\,\,{\bf{F}},J),
\end{aligned}
\end{equation}
\end{linenomath*}
$J=\det {\bf{F}}$, and the function ${{\bar \psi }^e}:{R^{19}} \to R$ is convex for all material points. 
\end{definition}
To construct constitutive equations that satisfy \textit{objectivity}, \textit{polyconvexity}, and \textit{material symmetry}, a further consideration is required. In this regard, we first recall a key sufficient statement from convex analysis: the composition $f(g(x))$  is convex in $x$ if $g$  is convex and $f$ is convex and non-decreasing. Now, based on this statement, one strategy is to express the Helmholtz free energy as a composition of a convex, non-decreasing function with a set of invariants of the deformation tensor that are chosen to ensure \textit{objectivity}, respect the material’s \textit{symmetry group}, and be convex with respect to input arguments of \eqref{b14}. By selecting these invariants appropriately, the overall free energy can be made to satisfy all three conditions simultaneously.   As shown in \cite{schroder2008anisotropic}, for second order structural tensors ${\mathbf{G}}_i, i=1,...,n$, one family of polyconvex invariants are  ${\text{I}}_i^{{\mathbf{G}},k} = {\left( {{\mathbf{C}}:{{\mathbf{G}}_i}} \right)^k},\, {\text{J}}_i^{{\mathbf{G}},k} = {\left( {{\text{cof}}\, \, {\mathbf{C}}:{{\mathbf{G}}_i}} \right)^k}$, $k \ge 1$ . In this study, we introduce the new data-driven framework in terms of  
\begin{linenomath*}
\begin{equation}\label{b15}
\begin{aligned}
{{\rm{I}}_{\bf{C}}}{\rm{ = }}{\bf{C}}:{\bf{I}},\,\,{\rm{I}}{{\rm{I}}_{\bf{C}}}{\rm{ = cof}}\,\,{\bf{C}}:{\bf{I}},\,\,{{\rm{I}}_i} = {\bf{C}}:{{\bf{L}}_i},\,\,{{\rm{J}}_i} = {\rm{cof}}\,\,{\bf{C}}:{{\bf{L}}_i}.
\end{aligned}
\end{equation}
\end{linenomath*}  
It is noted that  ${\rm{I}}_i^{{\bf{L}},k}$ and $ {{\rm{I}}_i}$ are convex with respect to $\bf{F}$ and ${\rm{J}}_i^{{\bf{L}},k}$ and $ {{\rm{J}}_i}$ are convex with respect to ${\rm{cof }}\,\,\,{\bf{F}}$. 

% For incompressible materials, we use the simplified version of the above invariants:
% \begin{linenomath*}
% \begin{equation}\label{b15-1}
% \begin{aligned}
% {{\rm{I}}_{\bf{C}}}={\bf{C}}:{\bf{I}},\,\,{\rm{I}}{{\rm{I}}_{\bf{C}}}={\bf{C}}^{-1}:{\bf{I}},\,\,{{\rm{I}}_i} = {\bf{C}}:{{\bf{L}}_i},\,\,{{\rm{J}}_i} ={\bf{C}}^{-1}:{{\bf{L}}_i}.
% \end{aligned}
% \end{equation}
% \end{linenomath*}
\subsection{Normality condition}
In formulating constitutive equations, we assume that when the Green–Cauchy strain tensor is equal to the identity tensor, i.e., ${{\bf{C}} = {\bf{I}}}$, the material is in a stress-free and energy-free state, irrespective of the state of internal variables. This assumption is referred to as the normality condition in \cite{linden2023neural} for hyper-elastic materials. Hence, the normality condition has the following form:   
\begin{linenomath*}
\begin{equation}\label{b15-2}
\begin{aligned}
&{\left. {{{\bar \psi }^e}} \right|_{{\mathbf{C}} = {\mathbf{I}}}} = 0,\\
&{\bf{S}} = 2{\left. {\frac{{\partial {{\bar \psi }^e}}}{{\partial {\bf{C}}}}} \right|_{{\bf{C}} = {\bf{I}}}} = {\bf{0}}.
\end{aligned}
\end{equation}
\end{linenomath*} 
It is worth mentioning that from  \eqref{b15} that all anisotropic invariants are equal to unity in the initial configuration, while the isotropic ones are equal to three.
\subsection{Growth condition}
Polyconvexity, along with the growth condition, secures the existence of a global minimizer. This condition states that the energy blows up to infinity for the case of extreme deformations. There are several ways to define growth condition (see, for example, \cite{ball1976convexity,muller1994new,holzapfel2000nonlinear}). For the case of data-driven methods, for hyper-elastic materials, one can add analytical terms to fulfill the growth condition analogous to \cite{klein2022polyconvex, linden2023neural}.  However, since such extreme modes of deformation rarely occur in real-world scenarios—and the numerical examples that follow lie well outside this regime—they remain primarily of theoretical interest, as also noted in \cite{klein2022polyconvex}. Therefore, such terms are not included in this study for simplicity. 
% One way to achieve this goal for compressible materials is the growth condition: \cite{holzapfel2000nonlinear}:
% \begin{linenomath*}
% \begin{equation}\label{b16}
% \begin{aligned}
% \mathop {\lim }\limits_{J \to {0^ + } } {{\tilde \psi }^e} =  + \infty ,\,\,\,\,\,\,\mathop {\lim }\limits_{J \to  + \infty } \tilde \psi  =  + \infty. 
% \end{aligned}
% \end{equation}
% \end{linenomath*} 
\subsection{Non-negativity of energy}
Finally, we assume that  Helmholtz potential energy ${{\bar \psi }^{e }} \ge 0$ for every admissible deformation, i.e., a deformation with $J> 0$. As mentioned in \cite{linden2023neural}, this condition is difficult to verify even for analytical models. For example, the non-negativity of the Neo-Hookean strain energy potential was recently addressed in \cite{linden2023neural}. Hence, verifying this condition in data-driven models is even more challenging, as these models have a generic structure, and—as we shall see—additional terms are required to satisfy the normality condition. Therefore, in this work, we numerically demonstrate that the trained energy density potentials remain non-negative in all numerical examples.     
\section{Damage framework}\label{damage_F}
In this part, the main architecture of the new constitutive equation is introduced, and how the aforementioned conditions are satisfied in the new framework is discussed. Specifically, we shall show that the new data-driven approach satisfies objectivity, thermodynamics consistency, material symmetry, and normality by construction. As discussed earlier, it is, in general, a daunting task to check for non-negativeness of the Helmholtz energy for all admissible deformations, even for analytical models  \cite{linden2023neural}. Hence, we show this property numerically, that for the available data, the strain energy density potentials remain  non-negative.
\subsection{Helmholtz free energy function}
To establish the new framework, the items described earlier are considered. As we model the nonlinear elasticity that incorporates damage,  the following 
Helmholtz energy function is considered:
\begin{linenomath*}
\begin{equation}\label{D0}
\begin{aligned}
{{\bar \psi }^{e{(aniso)} }} = {{\bar \psi }^{e{(aniso)} }}({\bf{C}},{\bf{D}}^{iso}, {\bf{D}}^{aniso}),
\end{aligned}
\end{equation}
\end{linenomath*} 
where ${\bf{D}}^{iso}=\alpha_0 \mathbf{I}$ with $\mathbf{I}$ is the second-order identity tensor, and ${\bf{D}}^{aniso}$ is the symmetric damage tensor for incorporation of the anisotropic damage. In this work, for simplicity, we establish the framework under the assumption that the principal directions of the damage tensor and those of the material symmetry in virgin material, if they exist, are co-axial:
\begin{linenomath*}
\begin{equation}\label{D1}
\begin{aligned}
&{{\mathbf{D}}^{aniso}} = \sum\limits_{i = 1}^3 {{\alpha _i}{{\mathbf{L}}_i}}  ,\\
&\mathcal{G} \equiv \left\{ {{\mathbf{Q}} \in {\text{O}}(3),\,\,{\mathbf{Q}}\,{{\mathbf{L}}_i}\,{{\mathbf{Q}}^T} = {{\mathbf{L}}_i}} \right\},\,\,\,i = 1,2,3. 
\end{aligned}
\end{equation}
\end{linenomath*} 
% where
% \begin{linenomath*}
% \begin{equation}\label{D1-1}
% \begin{aligned}
% & {{\mathbf{D}}^{iso}} = {\alpha _0}{\mathbf{I}},\\
% & {{\mathbf{D}}^{aniso}} = \sum\limits_{i = 1}^3 {{\alpha _i}{{\mathbf{L}}_i}}. 
% \end{aligned}
% \end{equation}
% \end{linenomath*} 
Also, it is further presumed that the damage principle directions remain the same during damage progress. Now, considering \eqref{D1} and \eqref{b13}, one can rewrite \eqref{D0} as:
\begin{linenomath*}
\begin{equation}\label{D2}
\begin{aligned}
{{\bar \psi }^{e(aniso)}} = {{\bar \psi }^e}({\mathbf{C}},{{\mathbf{L}}_i},{\alpha _i},{\alpha _0}),\,\,\,i = 1,2,3. 
\end{aligned}
\end{equation}
\end{linenomath*}
\begin{remark}
According to \eqref{b13}, the generalization for the non-alignment of the principal directions of the damage tensor and those of the material principles is straightforward. In this regard, by denoting ${{\bf{D}}^{iso}} = {\alpha _0}{\bf{I}}$ and ${{\bf{D}}^{aniso}} = \sum\limits_{i = 1}^3 {{\alpha _i}{\bf{n}}_i^{\bf{D}} \otimes {\bf{n}}_i^{\bf{D}}} $ with ${{\bf{n}}_i^{\bf{D}} \otimes {\bf{n}}_i^{\bf{D}}}, i=1,2,3$ as the principal direction of the damage tensor and ${\bf{n}}_i^o \otimes {\bf{n}}_i^o,\,\, i = 1,2,3$ as the structural tensors characterizing the orthotropic behavior of the virgin material, one can write:
\begin{linenomath*}
\begin{equation}\label{D3}
\begin{aligned}
{{\bar \psi }^{e(aniso)}} = {{\bar \psi }^e}({\bf{C}},{{\bf{L}}_j},{\bf{L}}_i^{\bf{D}},{\alpha _i},\alpha_0),\,\,\,\,{\bf{L}}_i^{\bf{D}} = {\bf{n}}_i^{\bf{D}} \otimes {\bf{n}}_i^{\bf{D}},{{\bf{L}}_j} = {\bf{n}}_j^o \otimes {\bf{n}}_j^o,\,\,i,j = 1,2,3.
\end{aligned}
\end{equation}
\end{linenomath*}
However, for simplicity, in this study, we focus on the constitutive equations arising from \eqref{D2}. 
\end{remark}
Now, by following \cite{itskov2004class}, we define generalized structural tensors
\begin{linenomath*}
\begin{equation}\label{D4}
\begin{aligned}
{{{\bf{\tilde L}}}^{(i)}} = \sum\limits_{j = 1}^3 {w_j^{(i)}{{\bf{L}}_j}} ,\,\,\, {\rm{tr}}({{{\bf{\tilde L}}}^{(i)}}) = 1,\,\,\,w_j^{\left( i \right)} \ge 0,\,\,\, i,j=1,2,3, 
\end{aligned}
\end{equation}
\end{linenomath*} 
% with the exclusion of the isotropic case 
% \begin{linenomath*}
% \begin{equation}\label{D4_1}
% \begin{aligned}
% {{{\bf{\tilde L}}}^{({i_L})}} \ne {\bf{I}},\,\, i_L=1,2,3,
% \end{aligned}
% \end{equation}
% \end{linenomath*} 
 where $\eqref{D4}_2$ can be imposed by the following:
\begin{linenomath*}
\begin{equation}\label{D5}
\begin{aligned}
\sum\limits_{j = 1}^3 {w_j^{\left( i \right)}}  = 1,\,i = 1,2,3.
\end{aligned}
\end{equation}
\end{linenomath*} 
As discussed earlier, the isotropic function ${{\bar \psi }^e}$ in \eqref{D2} is written in terms of invariants defined in \eqref{b15} and generalized structural tensors \eqref{D4}:
\begin{linenomath*}
\begin{equation}\label{D6}
\begin{aligned}
&{{\bar \psi }^e} = {{\bar \psi }^e}\left( {{{\tilde I}^{\left( i \right)}},{{\tilde J}^{\left( i \right)}},{\rm{II}}{{\rm{I}}_{\bf{C}}},{\alpha _i};{{\rm{I}}_{\bf{C}}},{\rm{I}}{{\rm{I}}_{\bf{C}}}} \right),\\
&{{\tilde I}^{\left( i \right)}} = {\rm{tr}}({\bf{C}}{{{\bf{\tilde L}}}^{(i)}}),\,\,{{\tilde J}^{\left( i \right)}} = {\rm{tr}}({\rm{cof}}({\bf{C}}){{{\bf{\tilde L}}}^{(i)}}),\,i = 1,2,3
\end{aligned}
\end{equation}
\end{linenomath*}
where the isotropic invariants ${\rm{I}}_{\bf{C}}$ and ${\rm{II}}_{\bf{C}}$ are added as auxiliary input variables. Next, similar to \cite{lemaitre1994mechanics},  it is assumed that ${{\bar \psi }^e}$ is convex with respect to the principal damage variables,  so-called generalized standard materials (see  \cite{maugin1999thermomechanics}). It is discussed in the sequel that this assumption leads to well-defined equations governing the evolution of damage variables $\alpha_0$ and $\alpha_i, i=1,2,3$.    Also, as damage has a weakening effect on the material's property, it is natural to consider  ${{\bar \psi }^e}$ is decreasing with the evolution of $\alpha_i,\, i=1,2,3$. Now, considering the above-mentioned properties,  similar to the additive decomposition equation  introduced in \cite{schroder2003invariant}, we suggest the following for compressible materials: 
\begin{linenomath*}
\begin{equation}\label{D7}
\begin{aligned}
{{\tilde \psi }^e} = {p_0}\left( {{\alpha _0}} \right){{\bar \psi }_0}({{\text{I}}_{\mathbf{C}}},{\text{I}}{{\text{I}}_{\mathbf{C}}},{\text{II}}{{\text{I}}_{\mathbf{C}}}) + \sum\limits_{i = 1}^3 {{p_i}\left( {{\alpha _i}} \right)\bar \psi _i^{\tilde I,\tilde J}({{\tilde I}^{\left( i \right)}},{{\tilde J}^{\left( i \right)}},{\text{II}}{{\text{I}}_{\mathbf{C}}};{{\text{I}}_{\mathbf{C}}},{\text{I}}{{\text{I}}_{\mathbf{C}}})}  ,
\end{aligned}
\end{equation}
\end{linenomath*}
% \begin{linenomath*}
% \begin{equation}\label{D7}
% \begin{aligned}
% {{\tilde \psi }^e} = \sum\limits_{i = 1}^3 {{p_i}\left( {{\alpha _i}} \right)\bar \psi _i^{\tilde I,\tilde J}({{\tilde I}^{\left( i \right)}},{{\tilde J}^{\left( i \right)}};{{\rm{I}}_{\bf{C}}},{\rm{I}}{{\rm{I}}_{\bf{C}}})}+ {{\bar \psi }^{vol}}\left( {{\rm{II}}{{\rm{I}}_{\bf{C}}}} \right),
% \end{aligned}
% \end{equation}
% \end{linenomath*}
where  ${p_0}\left( {{\alpha _0}} \right),\,\, {\rm{and}}\,\,{{p_i}\left( {{\alpha _i}} \right)},\,\, i=1,2,3$ are decreasing convex functions and $ {\bar \psi _0}$ and $\bar \psi _i^{\tilde I,\tilde J}$'s are non-decreasing convex functions of their input arguments, which together with invariants ${{\tilde I}^{\left( i \right)}},{{\tilde J}^{\left( i \right)}},{{\rm{I}}_{\bf{C}}},{\rm{I}}{{\rm{I}}_{\bf{C}}},\,{\rm{and}}\,{\rm{II}}{{\rm{I}}_{\bf{C}}}$ satisfy the polyconvexity condition.
\begin{remark}
It is noticed  that ${\rm{II}}{{\rm{I}}_{\bf{C}}}$ 
  is included in the energy functions in the above expression. As will be seen, this implies that damage evolution can also be influenced by volumetric deformation. While in most biological soft tissues and elastomers the predominant source of damage is deviatoric deformation, volumetric deformation can indeed induce damage in compressible and brittle materials such as rocks and concrete.       
\end{remark}
\begin{remark}
As shall be explained in the sequel, introduction of the generalized structural tensors in the formulation leads to defining the conjugate damage tensor in terms of invariants that are not necessarily aligned with the damage principal directions.    
\end{remark}
\begin{remark}
A generalization of \eqref{D7} can be constructed by letting $i$ in \eqref{D4} and \eqref{D5} be an integer with $i>3$. Then, a possible  generalization of \eqref{D7} can be formulated as:
\begin{linenomath*}
\begin{equation}\label{D7-1}
\begin{aligned}
&{{\tilde \psi }^e} = {p_0}\left( {{\alpha _0}} \right){{\bar \psi }_0}({{\textnormal{I}}_{\mathbf{C}}},{\textnormal{I}}{{\textnormal{I}}_{\mathbf{C}}},{\textnormal{II}}{{\textnormal{I}}_{\mathbf{C}}}) + \sum\limits_{i = 1}^3 {{p_i}\left( {{\alpha _i}} \right)\bar \psi _{i,j}^{\tilde I,\tilde J}({{\tilde I}^{\left( {{1_i}} \right)}},{{\tilde J}^{\left( {{1_i}} \right)}},...,{{\tilde I}^{\left( {{m_i}} \right)}},{{\tilde J}^{\left( {{m_i}} \right)}},...,\,{\textnormal{II}}{{\textnormal{I}}_{\mathbf{C}}};{{\textnormal{I}}_{\mathbf{C}}},{\textnormal{I}}{{\textnormal{I}}_{\mathbf{C}}})},\\
& {{\tilde I}^{\left( {{m_i}} \right)}} = {\textnormal{tr}}({\mathbf{C}}{{{\mathbf{\tilde L}}}^{\left( {{m_i}} \right)}}),\,\, {{\tilde J}^{\left( {{m_i}} \right)}} = {\textnormal{tr}}({\textnormal{cof}}({\mathbf{C}}){{{\mathbf{\tilde L}}}^{\left( {{m_i}} \right)}}),\\
& {{{\mathbf{\tilde L}}}^{({m_i})}} = \sum\limits_{j = 1}^3 {w_j^{({m_i})}{{\mathbf{L}}_j}} ,\,\,\, {\textnormal{tr}}({{{\mathbf{\tilde L}}}^{({m_i})}}) = 1,\,\,w_j^{\left( {{m_i}} \right)} \geqslant 0,\,\sum\limits_{j = 1}^3 {w_j^{\left( {{m_i}} \right)}}  = 1.
\end{aligned}
\end{equation}
\end{linenomath*}
\end{remark}
\subsection{Damage evolution}\label{damage_evol}
To complete the constitutive equations, it is necessary to define equations governing the evolution of $\alpha_0$ and $\alpha_i$. It is noticed that governing equations must be consistent with \eqref{b5}. A most common approach that is followed in associated plasticity is employing the principle of maximum dissipation and defining a convex dissipative potential function that governs the evolution of internal variables. This approach was used for damage in  \cite{steinmann1998framework, menzel2001theoretical}. In this work, we follow a similar concept and generalize the strain-based damage introduced for nonlinear isotropic and linear anisotropic materials in \cite{simo1987strain}. Let us denote $y_0$ and $y_i$, respectively,  as the conjugate thermodynamic force corresponding to damage variables $\alpha_0$ and  $\alpha_i$, which according to the representation \eqref{D7}, can be calculated from:
\begin{linenomath*}
\begin{equation}\label{D8}
\begin{aligned}
& {y_0} =  - \frac{{d{p_0}}}{{d{\alpha _0}}}{{\bar \psi }_0},\\
& {y_i} =  - \frac{{d{p_i}}}{{d{\alpha _i}}}\bar \psi _i^{\tilde I,\tilde J},\,\,\,i = 1,2,3. 
\end{aligned}
\end{equation}
\end{linenomath*}
Clearly, due to the definition of the generalized structural tensors, $y_i$ is affected by invariants that are calculated from the inner product of strain metric with ${{{\bf{\tilde L}}}^{(i_L)}}$ that are not necessarily aligned with principal direction ${\mathbf{L}}_i$.  By postulating that the evolution for $\alpha_i$ ($\alpha_0$) is governed through the corresponding  $y_i$ ($y_0$) and introducing $r_i$ ($r_0$) as the corresponding  damage threshold, one can define the following damage potential:
\begin{linenomath*}
\begin{equation}\label{D9}
\begin{aligned}
&g = g_0(y_0,r_0)+\sum\limits_{i = 1}^3 {{g_i}({y_i},{r_i})},\\
& g_0({y_0},{r_0})=G_0(y_0)-G_0(r_0),\\
& {g_i}({y_i},{r_i}) = {G_i}({y_i}) - {G_i}({r_i}),\,\,\,\,i = 1,2,3,
\end{aligned}
\end{equation}
\end{linenomath*}
where $G_0$ and $G_i$'s are increasing functions. \\
% with the condition:
% \begin{linenomath*}
% \begin{equation}\label{D10}
% \begin{aligned}
% \mathop {\lim }\limits_{y \to \infty } \frac{{d{G_i}\left( y \right)}}{{dy}} = 0, i=1,2,3. 
% \end{aligned}
% \end{equation}
% \end{linenomath*}
Now, analogous to \cite{simo1987strain}, we define the damage evolution equations as:
\begin{linenomath*}
\begin{equation}\label{D12}
\begin{aligned}
&{\dot \alpha _0} = {\dot \mu _0}\frac{{\partial {g_0}}}{{\partial {y_0}}} = {\dot \mu _0}\frac{{d{G_0}}}{{d{y_0}}},\\
&{{\dot r}_0}={{\dot \mu}_0},\\
&{{\dot \alpha }_i} = {{\dot \mu }_i}\frac{{\partial {g_i}}}{{\partial {y_i}}} = {{\dot \mu }_i}\frac{{d{G_i}}}{{d{y_i}}},\\
&{{\dot r}_i}={{\dot \mu}_i},\,\, i=1,2,3,
\end{aligned}
\end{equation}
\end{linenomath*}
with $\mu_i$'s being multipliers yet to be determined, referred to as damage consistency parameters. Next, we make sure evolution in damage in anisotropic (isotropic) case does not occur when $g_i<0$ ($g_0<0$), and when it occurs $g_i=0$ ($g_0=0$), namely, the consistency condition. Furthermore, to account for loading and unloading conditions, we consider the KKT condition as follows:
\begin{linenomath*}
\begin{equation}\label{D13}
\begin{aligned}
&{\dot \mu _0} \ge 0,\,\,\,\,\, g_0(y_0,r_0) \le 0,\,\,\,\,\, \dot \mu_0 g_0=0, \\
&{\dot \mu _i} \ge 0,\,\,\,\,\, g_i(y_i,r_i) \le 0,\,\,\,\,\, \dot \mu_i g_i=0 .
\end{aligned}
\end{equation}
\end{linenomath*}
In addition to that, we have the consistency condition by which  $G_i=0$ ($G_0=0$) during damage evolution, which along with increasing property of $G_i$ ($G_0$) imply the following:
\begin{linenomath*}
\begin{equation}\label{D14}
\begin{aligned}
&{{\dot \mu}_0} = {{\dot y}_0},\,\,\,\, {r_0}={y_0},\\
&{{\dot \mu}_i} = {{\dot y}_i},\,\,\,\, {r_i}={y_i},\,\, i=1,2,3.
\end{aligned}
\end{equation}
\end{linenomath*}
Now, by differentiating \eqref{D8} and making use of \eqref{D14}, one can rearrange \eqref{D12} as follows during damage:
\begin{linenomath*}
\begin{equation}\label{D15}
\begin{aligned}
&{{\dot \alpha }_0} = \frac{{\left[ { - \frac{{d{G_0}}}{{d{y_0}}}\frac{{d{p_0}}}{{d{\alpha _0}}}\left( {\frac{{d{{\bar \psi }_0}}}{{d{\mathbf{C}}}}:{\mathbf{\dot C}}} \right)} \right]}}{{1 + \frac{{d{G_0}}}{{d{y_0}}}\frac{{{d^2}{p_0}}}{{d\alpha _0^2}}{{\bar \psi }_0}}},\\
&{{\dot \alpha }_i} = \frac{{\left[ { - \frac{{d{G_i}}}{{d{y_i}}}\frac{{d{p_i}}}{{d{\alpha _i}}}\left( {\frac{{d\bar \psi _i^{I,J}}}{{d{\mathbf{C}}}}:{\mathbf{\dot C}}} \right)} \right]}}{{1 + \frac{{d{G_i}}}{{d{y_i}}}\frac{{{d^2}{p_i}}}{{d\alpha _i^2}}\bar \psi _i^{I,J}}},\,\,\,\, i = 1,2,3.
\end{aligned}
\end{equation}
\end{linenomath*}
\begin{remark}
As can be seen from the increasing nature of $G_i$ ($G_0$) and the convexity of the attenuation function $p_i$ ($p_0$), the right-hand side of  \eqref{D15} is well defined, as the denominator remains strictly positive. 
\end{remark}
\begin{remark}
It is observed that the governing evolution equations are consistent with the C-D condition $\eqref{b5}_2$. In fact, from $\eqref{D12}$, the monotonicity of $G_i$ and $G_0$, and the KKT condition $\eqref{D13}$, one can conclude that:
\begin{linenomath*}
\begin{equation}\label{D16}
\begin{aligned}
&{y_0}{{\dot \alpha }_0} \ge 0,\\
&{y_i}{{\dot \alpha }_i} \ge 0,\,\,\, i=1,2,3.
\end{aligned}
\end{equation}
\end{linenomath*}
\end{remark}
\section{Anisotropy induced by damage }\label{induced_Aniso}
The above formulation represents a damage framework developed for orthotropic materials. In this part, we specialize this framework for the case where the virgin material is isotropic, and anisotropy emerges due to damage.\\
In this regard,  the initial isotropy implies that the evolution laws for different directions should be the same for all directions, leading to $G_i=G, i=1,2,3$ and the same attenuation functions $P_i, i=1,2,3$. Moreover, when material is under biaxial or three-axial loadings, the stresses in the loading directions must be the same. This leads to  ${\bar \psi _i^{I,J}}=\bar \psi, i=1,2,3$:
\begin{linenomath*}
\begin{equation}\label{D17}
\begin{aligned}
{{{\bf{\tilde L}}}^{(i)}} = {{\bf{L}}_i},\,\,{{\tilde I}^{\left( i \right)}} = {I_i},\,\,{{\tilde J}^{\left( i \right)}} = {J_i},\,\,\,i = 1,2,3. 
\end{aligned}
\end{equation}
\end{linenomath*}
Hence, for anisotropy induced by damage, we propose the following generic Helmholtz form:
\begin{linenomath*}
\begin{equation}\label{D18}
\begin{aligned}
{{\tilde \psi }^e} = {p_0}\left( {{\alpha _0}} \right){{\bar \psi }_0}({{\text{I}}_{\mathbf{C}}},{\text{I}}{{\text{I}}_{\mathbf{C}}},{\text{II}}{{\text{I}}_{\mathbf{C}}}) + \sum\limits_{i = 1}^3 {p\left( {{\alpha _i}} \right)\bar \psi ({{ I}^{\left( i \right)}},{{ J}^{\left( i \right)}},{\text{II}}{{\text{I}}_{\mathbf{C}}};{{\text{I}}_{\mathbf{C}}},{\text{I}}{{\text{I}}_{\mathbf{C}}})}.
\end{aligned}
\end{equation}
\end{linenomath*}
Subsequently, the damage evolution can be represented as:
\begin{linenomath*}
\begin{equation}\label{D19}
\begin{aligned}
&{{\dot \alpha }_0} = {{\dot \mu }_0}\frac{{d{G_0}}}{{d{y_0}}}\,\,,\,\,{{\dot \mu }_0} > 0,\,{{\dot \mu }_0}{G_0} = 0,\\
&{{\dot \alpha }_i} = {{\dot \mu }_i}\frac{{dG}}{{d{y_i}}}\,\,,\,\,{{\dot \mu }_i} > 0,\,{{\dot \mu }_i}G = 0,\,\,\,i = 1,2,3,
\end{aligned}
\end{equation}
\end{linenomath*}
leading to the following equation for the anisotropic part during damage evolution:
\begin{linenomath*}
\begin{equation}\label{D20}
\begin{aligned}
{{\dot \alpha }_i} = \frac{{\left[ { - \frac{{dG}}{{d{y_i}}}\frac{{dp}}{{d{\alpha _i}}}\left( {\frac{{d\bar \psi }}{{d{\mathbf{C}}}}:{\mathbf{\dot C}}} \right)} \right]}}{{1 + \frac{{dG}}{{d{y_i}}}\frac{{{d^2}p}}{{d\alpha _i^2}}\bar \psi }},\,\,\, i=1,2,3, 
\end{aligned}
\end{equation}
and $\eqref{D15}_1$ remains the same. 
\end{linenomath*}
\section{Data-driven framework}
In this section, we introduce the data-driven framework that fulfills the items described in the background \ref{back} and damage framework \ref{damage_F} by construction. The only exception is the non-negativity of the energy, which is challenging even for analytical constitutive models \cite{linden2023neural}. The numerical results demonstrate that the new model fulfills this condition. We define potentials ${\bar \psi _i^{\tilde I,\tilde J}}, i=1,2,3$ and  ${{\bar \psi }_0}$ in terms of \textit{a special architecture} of a multi-layer perceptrons (MLPs). In general an MLP consists of $k$ hidden layers with inputs and outputs denoted by vectors ${{\bf{z}}_{j - 1}} \in {\mathbb{R}^{{n_{j - 1}}}},\,\,\,{{\bf{z}}_j} \in {\mathbb{R}^{{n_j}}},\,\,\, j=1,...,k$, respectively, can be represented as:
\begin{linenomath*}
\begin{equation}\label{DD0}
\begin{aligned}
&{{\bf{z}}_0} = {\bf{x}},\\
&{{\bf{z}}_j} = \mathcal{A}_j\left( {{{\bf{W}}_j}{{\bf{z}}_{j - 1}} + {{\bf{b}}_j}} \right),\,\,\,j = 1,...,k,\,\,\,{{\bf{W}}_j} \in {\mathbb{R}^{{n_j} \times {n_{j - 1}}}},\,\,\,{{\bf{b}}_j} \in {\mathbb{R}^{{n_j}}},
\end{aligned}
\end{equation}
\end{linenomath*}
whose parameters are set of all  weights $\mathbb{W}$ and biases $\mathbb{B}$ and $\mathcal{A}_j$'s are activation functions. 
\subsection{Objectivity and material symmetry}
Since the inputs to the MLPs are invariants, the data-driven framework is inherently objective. Furthermore, the incorporation of structural tensors in the definition of these invariants accounts for material anisotropy, whether it is inherent in the virgin material or induced by damage. 
\subsection{Polyconvexity}
To ensure the polyconvexity condition by construction in the data-driven form, we use a specific format of MLPs, known as ICNN architecture \cite{amos2017input}, analogous to \cite{klein2022polyconvex, linden2023neural}. It is noted that to reduce the computational cost, the ICNNs we use in the present study do not have skipping layers as proposed in \cite{amos2017input}. To proceed,  it is sufficient to consider ${{\bar \psi }^{0}}$ and $\tilde\psi_i^ {I,J}$'s as non-decreasing and convex functions of their input arguments (see \cite{amos2017input, klein2022polyconvex, linden2023neural}), given that the inputs are a vector of the polyconvex invariants.  Now, considering ${\bf{x}} = {\left[ {{{\tilde I}^{\left( i \right)}},\,{{\tilde J}^{\left( i \right)}},\,{{\rm{I}}_{\bf{C}}},\,{\rm{I}}{{\rm{I}}_{\bf{C}}},\,{\rm{II}}{{\rm{I}}_{\bf{C}}}} \right]^T},\,i = 1,2,3$, to guarantee polyconvexity, we have the following conditions \cite{klein2022polyconvex, linden2023neural}:
\begin{linenomath*}
\begin{equation}\label{DD1}
\begin{aligned}
& \mathcal{A}_j\,\,\, \rm{convex\,\,\, and\,\,\, non-decreasing},\\
& {\mathbf{W}}_j \in \mathbb{R}^{n_j \times n_{j-1}}_{\ge 0},\,\,\, j=1,...,m,\\
& {{\mathbf{b}}_j} \in {\mathbb{R}^{{n_j}}},\,\,\, j = 1,...,m.
\end{aligned}
\end{equation}
\end{linenomath*} 
Hence, in the new data-driven framework, we represent ${{\bar \psi }_0}$ and ${\bar \psi _i^{\tilde I,\tilde J}},\ i = 1, 2, 3$, using the above-defined ICNN architecture, scaled by a constant positive factor. These representations are denoted by ${{\bar \psi }_0^{({ICNN})}}$ and ${\bar \psi _i^{({ICNN})}},\ i = 1, 2, 3$.
\subsection{Thermodynamic consistency}
 To make the method thermodynamically consistent  in the sense of the definition \ref{th_consistency}, we express the stress as:
\begin{linenomath*}
\begin{equation}\label{DD2}
\begin{aligned}
{\mathbf{S}} =&2{p_0}\left( {{\alpha _0}} \right)\left( {\frac{{\partial \hat \psi _0^{(ICNN)}}}{{\partial {{\rm{I}}_{\bf{C}}}}}{\bf{I}} + \frac{{\partial \hat \psi _0^{(ICNN)}}}{{\partial {\rm{I}}{{\rm{I}}_{\bf{C}}}}}\left( {{{\rm{I}}_{\bf{C}}}{\bf{I}} - {\bf{C}}} \right) + \frac{{\partial \hat \psi _0^{(ICNN)}}}{{\partial {\rm{II}}{{\rm{I}}_{\bf{C}}}}}\left( {{\rm{II}}{{\rm{I}}_{\bf{C}}}{\bf{C}}^{-1}} \right)} \right) +\\
& 2\sum\limits_{i = 1}^3 {{p_i}\left( {{\alpha _i}} \right)\left( {\frac{{\partial \hat \psi _i^{(ICNN)}}}{{\partial {{\tilde I}^{\left( i \right)}}}}{{{\bf{\tilde L}}}^{(r)}} + \frac{{\partial \hat \psi _i^{(ICNN)}}}{{\partial {{\tilde J}^{\left( i \right)}}}}\left( {{{\tilde J}^{\left( i \right)}}{\bf{I}} - {\rm{II}}{{\rm{I}}_{\bf{C}}}{{\bf{C}}^{ - 1}}{{{\bf{\tilde L}}}^{(r)}}} \right){{\bf{C}}^{ - 1}}} \right)} +\\
& 2\sum\limits_{i = 1}^3 {{p_i}\left( {{\alpha _i}} \right)\left( {\frac{{\partial \hat \psi _i^{(ICNN)}}}{{\partial {{\rm{I}}_{\bf{C}}}}}{\bf{I}} + \frac{{\partial \hat \psi _i^{(ICNN)}}}{{\partial {\rm{I}}{{\rm{I}}_{\bf{C}}}}}\left( {{{\rm{I}}_{\bf{C}}}{\bf{I}} - {\bf{C}}} \right) + \frac{{\partial \hat \psi _i^{(ICNN)}}}{{\partial {\rm{II}}{{\rm{I}}_{\bf{C}}}}}\left( {{\rm{II}}{{\rm{I}}_{\bf{C}}}{\bf{C}}^{-1}} \right)} \right)}+2\frac{{\partial {\psi ^{stress}}}}{{\partial {\bf{C}}}},
\end{aligned}
\end{equation}
\end{linenomath*}
where $\hat \psi_0^{({ICNN})}$, $\hat \psi_i^{({ICNN})}$, and $\psi^{{stress}}$ are yet to be defined under the normality conditions. The above expression satisfies the symmetry condition ${\bf{S}}={\bf{S}}^T$. Similarly, the damage conjugate force tensors are defined as  
\begin{linenomath*}
\begin{equation}\label{DD3}
\begin{aligned}
&{{\bf{y}}^{iso}} =  - \frac{{\partial {{\tilde \psi }^e}}}{{\partial {{\bf{D}}^{iso}}}} = \underbrace { - \frac{{d{p_0}}}{{d{\alpha _0}}}\hat \psi _0^{  (ICNN)} - \frac{{d{\psi ^{stress}}}}{{d{\alpha _0}}}}_{{y_0}}{\bf{I}}\\
&{{\bf{y}}^{aniso}} =  - \frac{{\partial {{\tilde \psi }^e}}}{{\partial {{\bf{D}}^{aniso}}}} = \sum\limits_{i = 1}^3 {\underbrace { - \frac{{d{p_i}}}{{d{\alpha _i}}}\hat \psi _i^{(ICNN)} - \frac{{d{\psi ^{stress}}}}{{d{\alpha _i}}}}_{{y_i}}{{\bf{L}}_i}},
\end{aligned}
\end{equation}
\end{linenomath*} 
Furthermore, to  construct  decreasing and convex $p_i$'s and $p_0$, we employ a weighted summation of nonlinear functions. In particular,  we note that the family of functions ${{{\left( {1 - \frac{{{\alpha}}}{{a}}} \right)}^{{q_j}}}}$ with $q_j  \ge 1$ and $a > 0$  are convex decreasing function in range $0 \le \alpha\le a$. Hence, one can define a generic format as follows:
\begin{linenomath*}
\begin{equation}\label{DD4}
\begin{aligned}
{p}({\alpha};{\bf{w}},{a}) =&\left\{ \begin{array}{l}
\sum\limits_j {{w_j}{{\left( {1 - \frac{{{\alpha }}}{{a}}} \right)}^{{q_j}}}} \,\,\,\,\,\,\,\,\,\,\,0 \le {\alpha} \le a\\
0\,\,\,\,\,\,\,\,\,\,\,\,\,\,\,\,\,\,\,\,\,\,\,\,\,\,\,\,\,\,\,\,\,\,\,\,\,\,\,\,\,\,\,\,\,\,\,\,\,\,\,\,\,\,\,\,\,\,\,{\alpha} \ge a
\end{array} \right., \\
& \sum\limits_j {{w_j}}  = 1,
\end{aligned}
\end{equation}
\end{linenomath*}
where $q_j \ge 1$ are predefined values that remain fixed during training, and weights $w_j$ and cutoff coefficients $a^0_i$ are parameters to be tuned by training. $\eqref{DD4}_2$ is imposed due to ${p_i}(0;{\bf{w}},{a_0})=1$, which states that there is no deterioration in the material's property when the internal variables vanish. 
\begin{remark}
It is noticed that in the new framework, in contrast to classical $[1-d]$ approaches, the degradation due to damage is expressed as weighted summation of  non-linear functions where the damage parameters are no longer restricted to remain $ 0 \leqslant {\alpha _0},\, {\alpha _i} \leqslant 1,\,\,\, i = 1,2,3$.   
\end{remark}
Finally, to generate increasing damage potential, we noticed that $G_i$'s are used in terms of their derivative in the damage evolution equations \eqref{D12}. Thus, we define
\begin{linenomath*}
\begin{equation}\label{DD5}
\begin{aligned}
&\frac{{d{G_0}}}{{d{y_0}}} = {{G'}_0}^{\,(MLP)},\\
&\frac{{d{G_i}}}{{d{y_i}}} = {{G'}_i}^{\,(MLP)},\,\,\,i = 1,2,3,   
\end{aligned}
\end{equation}
\end{linenomath*}
with condition ${{G'}_0}^{(MLP)}({y_0}) \ge 0$ and ${{G'}_i}^{(MLP)}({y_i}) \ge 0$ for $y_0 \ge 0$ and $y_i \ge 0,\ i = 1, 2, 3$. To ensure the aforementioned non-negativity condition, in this work, we use the following:

\begin{linenomath*}
\begin{equation}\label{DD5-1}
\begin{aligned}
&{{G'}_0}^{(MLP)}\left( {{y_0};\,{\Theta _{{G_0}}},{c_0}} \right) = \exp \left( { - {c_0}{y_0}} \right){\rm{softplus}}\left( {{\rm{MLP}}\left( {{y_0};{\Theta _{{G_0}}}} \right)} \right),\,\,\,{y_0} \ge 0\\
&{G'_i}^{\,\,(MLP)}\left( {{y_i};\,{\Theta _{{G_i}}},{c_i}} \right) = \exp \left( { - {c_i}{y_i}} \right){\rm{softplus}}\left( {{\rm{MLP}}\left( {{y_i};\,{\Theta _{{G_i}}}} \right)} \right),\,\,{y_i} \ge 0,\,\,\,i = 1,2,3,  
\end{aligned}
\end{equation}
\end{linenomath*}
in which  $c_0$ and $c_i$ are non-negative parameters. 
\subsection{Normality condition}
One significant condition is that the material is supposed to be energy and stress-free in the absence of loads. This condition is called normality of stress and energy. While this condition for analytical methods is more convenient because one can change it case by case, for data-driven methods, the existence of generic ICNN functions makes it more challenging. It is noted that, in data-driven methods, the stress-free condition is typically fulfilled approximately, as the stress–strain ground-truth curves fulfill this condition. However, it is desirable to develop a data-driven method that satisfy these conditions by construction.  In this section, analogous to \cite{linden2023neural}, we introduce additional terms to precisely satisfy the normality conditions. For the vanishing condition of  energy for compressible materials, one can add an extra term as follows:
% \begin{linenomath*}
% \begin{equation}\label{4-10}
% \begin{aligned}
% &{{\tilde \psi }^e} = \sum\limits_{i = 1}^s {{p_i}\left( {{\alpha _i}} \right)\tilde \psi _i^{I,J}({{\tilde I}^{\left( i \right)}},{{\tilde J}^{\left( i \right)}})}  + {{\tilde \psi }^{vol}}\left( {{\rm{II}}{{\rm{I}}_{\bf{C}}}} \right), s \in \mathbb{Z}^+\\
% &{\tilde I^{\left( i \right)}} = \sum\limits_{j = 1}^3 {w_j^{\left( i \right)}{I_j}} ,{\tilde J^{\left( i \right)}} = \sum\limits_{j = 1}^3 {w_j^{\left( i \right)}{J_j}} ,w_j^{\left( r \right)} \ge 0,\sum\limits_{j = 1}^3 {w_j^{\left( i \right)}}  = 1,\\
% & {I_j} = {\rm{tr}}({\bf{C}}{{\bf{L}}_j}), {J_j} = {\rm{tr}}({\rm{cof}}({\bf{C}}){{\bf{L}}_j})\\
% &{{{\bf{\tilde L}}}^{(r)}} = \sum\limits_{j = 1}^3 {w_j^{(r)}{{\bf{L}}_j}}, {\rm{tr}}({{{\bf{\tilde L}}}^{(r)}}) = 1
% \end{aligned}
% \end{equation}
% \end{linenomath*}
% Therefore, to satisfy the normality of energy, we consider:
\begin{linenomath*}
\begin{equation}\label{4-11}
\begin{aligned}
&\hat \psi _0^{(ICNN)}({{\text{I}}_{\mathbf{C}}},{\text{I}}{{\text{I}}_{\mathbf{C}}},{\text{II}}{{\text{I}}_{\mathbf{C}}}) = \bar \psi _0^{(ICNN)}({{\text{I}}_{\mathbf{C}}},{\text{I}}{{\text{I}}_{\mathbf{C}}},{\text{II}}{{\text{I}}_{\mathbf{C}}}) - \bar \psi _0^{(ICNN)}(3,3,1),\\
&\hat \psi _i^{(ICNN)}({{\tilde I}^{\left( i \right)}},{{\tilde J}^{\left( i \right)}};{{\text{I}}_{\mathbf{C}}},{\text{I}}{{\text{I}}_{\mathbf{C}}},{\text{II}}{{\text{I}}_{\mathbf{C}}}) = \bar \psi _i^{(ICNN)}({{\tilde I}^{\left( i \right)}},{{\tilde J}^{\left( i \right)}};{{\text{I}}_{\mathbf{C}}},{\text{I}}{{\text{I}}_{\mathbf{C}}},{\text{II}}{{\text{I}}_{\mathbf{C}}}) - \bar \psi _i^{(ICNN)}(1,1;3,3,1),\,\,\,i = 1,2,3.
\end{aligned}
\end{equation}
\end{linenomath*}
% The above condition is more natural than \eqref{4-4} because as far as I understood it, ${{{\tilde I}^{\left( i \right)}}}$ and ${{{\tilde J}^{\left( i \right)}}}$ are counterpart because one is based on  $\mathbf{C}$ and the other is based on $\mathbf{C}^{-1}$, and increase of the one is somehow accompanied by decrease of the other one or vice versa.  Thus, I think it would make more sense to define convex increasing  neural networks which are functions of  pairs of $\tilde{I}  ^{(i)}$ and $\tilde {J}^{(i)}$ , which accords to the strain energy function developed in    \cite{itskov2004class} . 
For the stress-free condition, on the other hand, the non-zero terms are:
\begin{linenomath*}
\begin{equation}\label{4-12}
\begin{aligned}
&{p_0}\left( {{\alpha _0}} \right)\underbrace {{{\left. {\left( {\frac{{\partial \hat \psi _0^{(ICNN)}}}{{\partial {{\rm{I}}_{\bf{C}}}}} + 2\frac{{\partial \hat \psi _0^{(ICNN)}}}{{\partial {\rm{I}}{{\rm{I}}_{\bf{C}}}}} + \frac{{\partial \hat \psi _0^{(ICNN)}}}{{\partial {\rm{II}}{{\rm{I}}_{\bf{C}}}}}} \right)} \right|}_{\left( {{{\rm{I}}_{\bf{C}}},\,{\rm{I}}{{\rm{I}}_{\bf{C}}},\,{\rm{II}}{{\rm{I}}_{\bf{C}}}} \right) = \left( {3,3,1} \right)}}}_{ \equiv R_0^{{{\rm{I}}_{\bf{C}}},{\rm{I}}{{\rm{I}}_{\bf{C}}},{\rm{II}}{{\rm{I}}_{\bf{C}}}}}{\bf{I}} +  \\
& \sum\limits_{i = 1}^3 {{p_i}\left( {{\alpha _i}} \right)\underbrace {{{\left. {\left( {\frac{{\partial \hat \psi _i^{(ICNN)}}}{{\partial {{\tilde I}^{\left( i \right)}}}} - \frac{{\partial \hat \psi _i^{(ICNN)}}}{{\partial {{\tilde J}^{\left( i \right)}}}}} \right)} \right|}_{\left( {{{\tilde I}^{\left( i \right)}},\,{{\tilde J}^{\left( i \right)}},\,{{\rm{I}}_{\bf{C}}},\,{\rm{I}}{{\rm{I}}_{\bf{C}}},\,{\rm{II}}{{\rm{I}}_{\bf{C}}}} \right) = \left( {1,1,3,3,1} \right)}}}_{ \equiv R_i^{\tilde I,\tilde J}}} {{{\bf{\tilde L}}}^{(i)}}+\\
& \sum\limits_{i = 1}^3 {{p_i}\left( {{\alpha _i}} \right)\underbrace {{{\left. {\left( {\frac{{\partial \hat \psi _i^{(ICNN)}}}{{\partial {{\tilde J}^{\left( i \right)}}}} + \frac{{\partial \hat \psi _i^{(ICNN)}}}{{\partial {{\rm{I}}_{\bf{C}}}}} + 2\frac{{\partial \hat \psi _i^{(ICNN)}}}{{\partial {\rm{I}}{{\rm{I}}_{\bf{C}}}}} + \frac{{\partial \hat \psi _i^{(ICNN)}}}{{\partial {\rm{II}}{{\rm{I}}_{\bf{C}}}}}} \right)} \right|}_{\left( {{{\tilde I}^{\left( i \right)}},\,{{\tilde J}^{\left( i \right)}},\,{{\rm{I}}_{\bf{C}}},\,{\rm{I}}{{\rm{I}}_{\bf{C}}},\,{\rm{II}}{{\rm{I}}_{\bf{C}}}} \right) = \left( {1,1,3,3,1} \right)}}}_{ \equiv R_i^{{{\rm{I}}_{\bf{C}}},{\rm{I}}{{\rm{I}}_{\bf{C}}},{\rm{II}}{{\rm{I}}_{\bf{C}}},\tilde J}}{\bf{I}}}    \ne {\bf{0}}.
\end{aligned}
\end{equation}
\end{linenomath*}
Now, we can write the counterpart as:
 \begin{linenomath*}
\begin{equation}\label{4-13}
\begin{aligned}
&- {p_0}\left( {{\alpha _0}} \right)R_0^{{{\rm{I}}_{\bf{C}}},{\rm{I}}{{\rm{I}}_{\bf{C}}},{\rm{II}}{{\rm{I}}_{\bf{C}}}}{\bf{I}} - \sum\limits_{i = 1}^3 {{p_i}\left( {{\alpha _i}} \right)R_i^{\tilde I,\tilde J}{{{\bf{\tilde L}}}^{(i)}}}  - \sum\limits_{i = 1}^3 {{p_i}\left( {{\alpha _i}} \right)R_i^{{{\rm{I}}_{\bf{C}}},\,{\rm{I}}{{\rm{I}}_{\bf{C}}},\,{\rm{II}}{{\rm{I}}_{\bf{C}}},\,\tilde J}{\bf{I}}}=  \\
&{\rm{ReLU}}\left( {{p_0}\left( {{\alpha _0}} \right)R_0^{{{\rm{I}}_{\bf{C}}},{\rm{I}}{{\rm{I}}_{\bf{C}}},{\rm{II}}{{\rm{I}}_{\bf{C}}}}} \right)\left( {{{\left. {\frac{{d{\rm{I}}{{\rm{I}}_{\bf{C}}}}}{{d{\bf{C}}}}} \right|}_{{\bf{C}} = {\bf{I}}}} - 3{\bf{I}}} \right) + {\rm{ReLU}}\left( {{\rm{ - }}{p_0}\left( {{\alpha _0}} \right)R_0^{{{\rm{I}}_{\bf{C}}},{\rm{I}}{{\rm{I}}_{\bf{C}}},{\rm{II}}{{\rm{I}}_{\bf{C}}}}} \right){\left. {\frac{{d{{\rm{I}}_{\bf{C}}}}}{{d{\bf{C}}}}} \right|_{{\bf{C}} = {\bf{I}}}}+ \\
&\sum\limits_{i = 1}^3 {{\rm{ReLU}}\left( {{p_i}\left( {{\alpha _i}} \right)R_i^{\tilde I,\tilde J}} \right)\left( {{{\left. {\frac{{d{{\tilde J}^{\left( i \right)}}}}{{d{\bf{C}}}}} \right|}_{{\bf{C}} = {\bf{I}}}} - {\bf{I}}} \right)}  + \sum\limits_{i = 1}^3 {{\rm{ReLU}}\left( {{\rm{ - }}{p_i}\left( {{\alpha _i}} \right)R_i^{\tilde I,\tilde J}} \right){{\left. {\frac{{d{{\tilde I}^{\left( i \right)}}}}{{d{\bf{C}}}}} \right|}_{{\bf{C}} = {\bf{I}}}} - \sum\limits_{i = 1}^3 {{p_i}\left( {{\alpha _i}} \right)R_i^{{{\rm{I}}_{\bf{C}}},{\rm{I}}{{\rm{I}}_{\bf{C}}},{\rm{II}}{{\rm{I}}_{\bf{C}}},\tilde J}{\bf{I}}} } ,
\end{aligned}
\end{equation}
\end{linenomath*}
with ${\rm{ReLU}}\left( x \right) = \max \left( {0,\,x} \right)$.
%, which can be further rearranged as
%  \begin{linenomath*}
% \begin{equation}\label{4-14}
% \begin{aligned}
% &{\rm{ReLU}}\left( {{p_0}\left( {{\alpha _0}} \right)R_0^{{{\rm{I}}_{\bf{C}}},{\rm{I}}{{\rm{I}}_{\bf{C}}},{\rm{II}}{{\rm{I}}_{\bf{C}}}}} \right){\left. {\frac{{d{\rm{I}}{{\rm{I}}_{\bf{C}}}}}{{d{\bf{C}}}}} \right|_{{\bf{C}} = {\bf{I}}}} + {\rm{ReLU}}\left( {{\rm{ - }}{p_0}\left( {{\alpha _0}} \right)R_0^{{{\rm{I}}_{\bf{C}}},{\rm{I}}{{\rm{I}}_{\bf{C}}},{\rm{II}}{{\rm{I}}_{\bf{C}}}}} \right){\left. {\frac{{d{{\rm{I}}_{\bf{C}}}}}{{d{\bf{C}}}}} \right|_{{\bf{C}} = {\bf{I}}}} +\\
% &\sum\limits_{i = 1}^3 {{\rm{ReLU}}\left( {{p_i}\left( {{\alpha _i}} \right)R_i^{\tilde I,\tilde J}} \right){{\left. {\frac{{d{{\tilde J}^{\left( i \right)}}}}{{d{\bf{C}}}}} \right|}_{{\bf{C}} = {\bf{I}}}}}  + \sum\limits_{i = 1}^3 {{\rm{ReLU}}\left( {{\rm{ - }}{p_i}\left( {{\alpha _i}} \right)R_i^{\tilde I,\tilde J}} \right){{\left. {\frac{{d{{\tilde I}^{\left( i \right)}}}}{{d{\bf{C}}}}} \right|}_{{\bf{C}} = {\bf{I}}}}}-\\
% &\left( {{\rm{3ReLU}}\left( {{p_0}\left( {{\alpha _0}} \right)R_0^{{{\rm{I}}_{\bf{C}}},{\rm{I}}{{\rm{I}}_{\bf{C}}},{\rm{II}}{{\rm{I}}_{\bf{C}}}}} \right) + \sum\limits_{i = 1}^3 {\left( {{\rm{ReLU}}\left( {{p_i}\left( {{\alpha _i}} \right)R_i^{\tilde I,\tilde J}} \right) + {p_i}\left( {{\alpha _i}} \right)R_i^{{{\rm{I}}_{\bf{C}}},{\rm{I}}{{\rm{I}}_{\bf{C}}},{\rm{II}}{{\rm{I}}_{\bf{C}}},\tilde J}} \right)} } \right){\bf{I}}.
% \end{aligned}
% \end{equation}
% \end{linenomath*}
Hence, we define:
\begin{linenomath*}
\begin{equation}\label{4-15}
\begin{aligned}
{\psi ^{{\rm{stress}}}} =& {\rm{ReLU}}({\rm{ - }}{p_0}\left( {{\alpha _0}} \right)R_0^{{{\rm{I}}_{\bf{C}}},{\rm{I}}{{\rm{I}}_{\bf{C}}},{\rm{II}}{{\rm{I}}_{\bf{C}}}})\left( {{{\rm{I}}_{\bf{C}}} - 3} \right) + {\rm{ReLU}}({p_0}\left( {{\alpha _0}} \right)R_0^{{{\rm{I}}_{\bf{C}}},{\rm{I}}{{\rm{I}}_{\bf{C}}},{\rm{II}}{{\rm{I}}_{\bf{C}}}})\left( {{\rm{I}}{{\rm{I}}_{\bf{C}}} - 3} \right)+ \\
&\sum\limits_{i = 1}^3 {{\rm{ReLU}}({\rm{ - }}{p_i}\left( {{\alpha _i}} \right)R_i^{\tilde I,\tilde J})\left( {{{\tilde I}^{\left( i \right)}} - 1} \right)}  + \sum\limits_{i = 1}^3 {{\rm{ReLU}}({p_i}\left( {{\alpha _i}} \right)R_i^{\tilde I,\tilde J})\left( {{{\tilde J}^{\left( i \right)}} - 1} \right)}   -\\
& 2\left( {{\rm{3ReLU}}({p_0}\left( {{\alpha _0}} \right)R_0^{{{\rm{I}}_{\bf{C}}},{\rm{I}}{{\rm{I}}_{\bf{C}}},{\rm{II}}{{\rm{I}}_{\bf{C}}}}) + \sum\limits_{i = 1}^3 {\left( {{\rm{ReLU}}({p_i}\left( {{\alpha _i}} \right)R_i^{\tilde I,\tilde J}) + {p_i}\left( {{\alpha _i}} \right)R_i^{{{\rm{I}}_{\bf{C}}},{\rm{I}}{{\rm{I}}_{\bf{C}}},{\rm{II}}{{\rm{I}}_{\bf{C}}},\tilde J}} \right)} } \right)(\sqrt {{\rm{II}}{{\rm{I}}_{\bf{C}}}}  - 1),
\end{aligned}
\end{equation}
\end{linenomath*}
for compressible materials, and therefore in what follows, we shall employ the following data-driven Helmholtz free energy function:
\begin{linenomath*}
\begin{equation}\label{4-15-1}
\begin{aligned}
{ \psi ^{\left( {DD} \right)}} = \tilde \psi _0^{(ICNN)} + \sum\limits_{i = 1}^3 {\tilde \psi _i^{(ICNN)}},
\end{aligned}
\end{equation}
\end{linenomath*}
with
\begin{linenomath*}
\begin{equation}\label{4-15-2}
\begin{aligned}
&\tilde \psi _0^{(ICNN)} = {p_0}\left( {{\alpha _0}} \right)\left[ {\hat \psi _0^{(ICNN)} + {\rm{ReLU}}\left( {{\rm{ - }}R_0^{{{\rm{I}}_{\bf{C}}},{\rm{I}}{{\rm{I}}_{\bf{C}}},{\rm{II}}{{\rm{I}}_{\bf{C}}}}} \right)\left( {{{\rm{I}}_{\bf{C}}} - 3} \right) + {\rm{ReLU}}\left( {R_0^{{{\rm{I}}_{\bf{C}}},{\rm{I}}{{\rm{I}}_{\bf{C}}},{\rm{II}}{{\rm{I}}_{\bf{C}}}}} \right)\left( {{\rm{I}}{{\rm{I}}_{\bf{C}}} - 3} \right)} \right]\\
 & - 6{p_0}\left( {{\alpha _0}} \right){\rm{ReLU}}\left( {R_0^{{{\rm{I}}_{\bf{C}}},{\rm{I}}{{\rm{I}}_{\bf{C}}},{\rm{II}}{{\rm{I}}_{\bf{C}}}}} \right)\left( {\sqrt {{\rm{II}}{{\rm{I}}_{\bf{C}}}}  - 1} \right),\\
&\tilde \psi _i^{(ICNN)} = {p_i}\left( {{\alpha _i}} \right)\left[ {\hat \psi _i^{(ICNN)} + \sum\limits_{i = 1}^3 {{\rm{ReLU}}\left( {{\rm{ - }}R_i^{\tilde I,\tilde J}} \right)\left( {{{\tilde I}^{\left( i \right)}} - 1} \right)}  + \sum\limits_{i = 1}^3 {{\rm{ReLU}}\left( {R_i^{\tilde I,\tilde J}} \right)\left( {{{\tilde J}^{\left( i \right)}} - 1} \right)} } \right] \\
&  - 2{p_i}\left( {{\alpha _i}} \right)\left( {{\rm{ReLU}}\left( {R_i^{\tilde I,\tilde J}} \right) + R_i^{{{\rm{I}}_{\bf{C}}},{\rm{I}}{{\rm{I}}_{\bf{C}}},{\rm{II}}{{\rm{I}}_{\bf{C}}},\tilde J}} \right)\left( {\sqrt {{\rm{II}}{{\rm{I}}_{\bf{C}}}}  - 1} \right).
\end{aligned}
\end{equation}
\end{linenomath*}
\begin{remark}
The above procedure to obtain $\psi^{\rm{stress}}$ shows that \eqref{4-15}, and consequently the forms defined in \eqref{4-15-1} \& \eqref{4-15-2}, are  not unique.  
\end{remark}
\section{Incompressible materials}
We have developed the new framework so far for compressible materials. However, there is a significant interest in enriching the literature with a data-driven damage formulation for incompressible materials, as incompressibility is a more realistic assumption for soft tissues and rubber-like materials. Hence, in this section, we explicitly present the modifications required to make the formulation applicable to incompressible materials.\\
First, it is noted that  the invariants defined \eqref{b15} are reduced to:
\begin{linenomath*}
\begin{equation}\label{inc-1}
\begin{aligned}
{{\text{I}}_{\mathbf{C}}}{\text{ = }}{\mathbf{C}}:{\mathbf{I}},\,  {\text{I}}{{\text{I}}_{\mathbf{C}}} =  {{\mathbf{C}}^{ - 1}}:{\mathbf{I}},\,  {{\text{I}}_i} = {\mathbf{C}}:{{\mathbf{L}}_i},\,  {{\text{J}}_i} =   {{\mathbf{C}}^{ - 1}}:{{\mathbf{L}}_i}.
\end{aligned}
\end{equation}
\end{linenomath*}
Next, \eqref{D7} is degenerated into:
\begin{linenomath*}
\begin{equation}\label{inc-2}
\begin{aligned}
{{\tilde \psi }^e} = {p_0}\left( {{\alpha _0}} \right){{\bar \psi }_0}\left( {{{\text{I}}_{\mathbf{C}}},{\text{I}}{{\text{I}}_{\mathbf{C}}}} \right) + \sum\limits_{i = 1}^3 {{p_i}\left( {{\alpha _i}} \right)\bar \psi _i^{\tilde I,\tilde J}\left( {{{\tilde I}^{\left( i \right)}},{{\tilde J}^{\left( i \right)}};{{\text{I}}_{\mathbf{C}}},{\text{I}}{{\text{I}}_{\mathbf{C}}}} \right)}  + {P_{pre}}\left( {\sqrt {{\text{II}}{{\text{I}}_{\mathbf{C}}}}  - 1} \right)
\end{aligned}
\end{equation}
\end{linenomath*}
where $P_{pre}$ is a Lagrange multiplier, namely the pressure,  determined from the solution of the elasticity boundary value problem augmented with the incompressibility condition. 
Since the damage evolutions are based on the damage variables and their conjugate forces, the theory elaborated in section \ref{damage_evol} remains the same for the case of the incompressible materials. similar to \eqref{D7},  \eqref{D18} is modified as:
\begin{linenomath*}
\begin{equation}\label{inc-3}
\begin{aligned}
{{\tilde \psi }^e} = {p_0}\left( {{\alpha _0}} \right){{\bar \psi }_0}\left( {{{\text{I}}_{\mathbf{C}}},{\text{I}}{{\text{I}}_{\mathbf{C}}}} \right) + \sum\limits_{i = 1}^3 {p\left( {{\alpha _i}} \right)\bar \psi \left( {{{ I}^{\left( i \right)}},{{ J}^{\left( i \right)}};{{\text{I}}_{\mathbf{C}}},{\text{I}}{{\text{I}}_{\mathbf{C}}}} \right)}  + {P_{pre}}\left( {\sqrt {{\text{II}}{{\text{I}}_{\mathbf{C}}}}  - 1} \right)
\end{aligned}
\end{equation}
\end{linenomath*}
for the case of anisotropy induced by damage for incompressible materials. Next, in incompressible materials, as the number of invariants and their definitions are slightly different than those for the compressible case, the stress is obtained from:
\begin{linenomath*}
\begin{equation}\label{inc-4}
\begin{aligned}
{\mathbf{S}} =& 2{p_0}\left( {{\alpha _0}} \right)\left( {\frac{{\partial \hat \psi _0^{(ICNN)}}}{{\partial {{\rm{I}}_{\bf{C}}}}}{\bf{I}} - \frac{{\partial \hat \psi _0^{(ICNN)}}}{{\partial {\rm{I}}{{\rm{I}}_{\bf{C}}}}}{{\left( {{{\bf{C}}^{ - 1}}} \right)}^2}} \right)+\\
& 2\sum\limits_{i = 1}^3 {{p_i}\left( {{\alpha _i}} \right)\left( {\frac{{\partial \hat \psi _i^{(ICNN)}}}{{\partial {{\tilde I}^{\left( i \right)}}}}{{{\bf{\tilde L}}}^{(r)}} - \frac{{\partial \hat \psi _i^{(ICNN)}}}{{\partial {{\tilde J}^{\left( i \right)}}}}{{\bf{C}}^{ - 1}}{{{\bf{\tilde L}}}^{(r)}}{{\bf{C}}^{ - 1}}} \right)}  + \\
& 2\sum\limits_{i = 1}^3 {{p_i}\left( {{\alpha _i}} \right)\left( {\frac{{\partial \hat \psi _i^{(ICNN)}}}{{\partial {{\rm{I}}_{\bf{C}}}}}{\bf{I}} - \frac{{\partial \hat \psi _i^{(ICNN)}}}{{\partial {\rm{I}}{{\rm{I}}_{\bf{C}}}}}{{\left( {{{\bf{C}}^{ - 1}}} \right)}^2}} \right)} +2\frac{{\partial {\psi ^{stress}}}}{{\partial {\bf{C}}}}+{P_{pr}}{{\bf{C}}^{ - 1}}.
\end{aligned}
\end{equation}
\end{linenomath*}
where $\hat \psi _0^{(ICNN)},\,\hat \psi _i^{(ICNN)}$, and ${\psi ^{stress}}$ shall be defined in what follows. In this regard, for incompressible material, the energy vanishing condition is obtained from:  
 \begin{linenomath*}
\begin{equation}\label{inc-5}
\begin{aligned}
&\hat \psi _0^{(ICNN)}({{\text{I}}_{\mathbf{C}}},{\text{I}}{{\text{I}}_{\mathbf{C}}}) = \bar\psi _0^{(ICNN)}({{\text{I}}_{\mathbf{C}}},{\text{I}}{{\text{I}}_{\mathbf{C}}}) - \bar \psi _0^{(ICNN)}(3,3)\\
&\hat \psi _i^{(ICNN)}({\tilde I^{\left( i \right)}},{\tilde J^{\left( i \right)}};{{\text{I}}_{\mathbf{C}}},{\text{I}}{{\text{I}}_{\mathbf{C}}}) = \bar \psi _i^{(ICNN)}({\tilde I^{\left( i \right)}},{\tilde J^{\left( i \right)}};{{\text{I}}_{\mathbf{C}}},{\text{I}}{{\text{I}}_{\mathbf{C}}}) - \bar \psi _i^{(ICNN)}(1,1;3,3)\,\,\,\,i = 1,2,3.
\end{aligned}
\end{equation}
\end{linenomath*}
For the normality condition for stress, the non-zero terms are:
\begin{linenomath*}
\begin{equation}\label{4-17}
\begin{aligned}
&{p_0}\left( {{\alpha _0}} \right)\underbrace {{{\left. {\left( {\frac{{\partial \hat \psi _0^{(ICNN)}}}{{\partial {{\text{I}}_{\mathbf{C}}}}} - \frac{{\partial \hat \psi _0^{(ICNN)}}}{{\partial {\text{I}}{{\text{I}}_{\mathbf{C}}}}}} \right)} \right|}_{\left( {{{\text{I}}_{\mathbf{C}}},{\kern 1pt} {\text{I}}{{\text{I}}_{\mathbf{C}}}} \right) = \left( {3,3} \right)}}}_{ \equiv R_0^{{{\text{I}}_{\mathbf{C}}},{\text{I}}{{\text{I}}_{\mathbf{C}}}}}{\mathbf{I}} + \sum\limits_{i = 1}^3 {{p_i}\left( {{\alpha _i}} \right)\underbrace {{{\left. {\left( {\frac{{\partial \hat \psi _i^{(ICNN)}}}{{\partial {{\tilde I}^{\left( i \right)}}}} - \frac{{\partial \hat \psi _i^{(ICNN)}}}{{\partial {{\tilde J}^{\left( i \right)}}}}} \right)} \right|}_{\left( {{{\tilde I}^{\left( i \right)}},{\kern 1pt} {{\tilde J}^{\left( i \right)}},{\kern 1pt} {{\text{I}}_{\mathbf{C}}},{\kern 1pt} {\text{I}}{{\text{I}}_{\mathbf{C}}}} \right) = \left( {1,1,3,3} \right)}}}_{ \equiv R_i^{\tilde I,\tilde J}}} {{\mathbf{\tilde L}}^{(i)}}  \\
& + \sum\limits_{i = 1}^3 {{p_i}\left( {{\alpha _i}} \right)\underbrace {{{\left. {\left( {\frac{{\partial \hat \psi _i^{(ICNN)}}}{{\partial {{\text{I}}_{\mathbf{C}}}}} - \frac{{\partial \hat \psi _i^{(ICNN)}}}{{\partial {\text{I}}{{\text{I}}_{\mathbf{C}}}}}} \right)} \right|}_{\left( {{{\tilde I}^{\left( i \right)}},{\kern 1pt} {{\tilde J}^{\left( i \right)}},{\kern 1pt} {{\text{I}}_{\mathbf{C}}},{\kern 1pt} {\text{I}}{{\text{I}}_{\mathbf{C}}}} \right) = \left( {1,1,3,3} \right)}}}_{ \equiv R_i^{{{\text{I}}_{\mathbf{C}}},{\text{I}}{{\text{I}}_{\mathbf{C}}}}}{\mathbf{I}}}   \ne {\bf{0}}.
\end{aligned}
\end{equation}
\end{linenomath*}
Now, analogous to procedure developed for the compressible materials earlier, one can define the following potential:
\begin{linenomath*}
\begin{equation}\label{4-18}
\begin{aligned}
{\psi ^{{\rm{stress}}}} =& {p_0}\left( {{\alpha _0}} \right)\left[ {{\rm{ReLU}}\left( { - R_0^{{{\rm{I}}_{\bf{C}}},{\rm{I}}{{\rm{I}}_{\bf{C}}}}} \right)\left( {{{\rm{I}}_{\bf{C}}} - 3} \right) + {\rm{ReLU}}\left( {R_0^{{{\rm{I}}_{\bf{C}}},{\rm{I}}{{\rm{I}}_{\bf{C}}}}} \right)\left( {{\rm{I}}{{\rm{I}}_{\bf{C}}} - 3} \right)} \right]+\\
&\sum\limits_{i = 1}^3 {{p_i}\left( {{\alpha _i}} \right)\left( {{\rm{ReLU}}\left( { - R_i^{{{\rm{I}}_{\bf{C}}},{\rm{I}}{{\rm{I}}_{\bf{C}}}}} \right)\left( {{{\rm{I}}_{\bf{C}}} - 3} \right) + {\rm{ReLU}}\left( {R_i^{{{\rm{I}}_{\bf{C}}},{\rm{I}}{{\rm{I}}_{\bf{C}}}}} \right)\left( {{\rm{I}}{{\rm{I}}_{\bf{C}}} - 3} \right)} \right)}+\\
&\sum\limits_{i = 1}^3 {{p_i}\left( {{\alpha _i}} \right)\left( {{\rm{ReLU}}\left( {{\rm{ - }}R_i^{\tilde I,\tilde J}} \right)\left( {{{\tilde I}^{\left( i \right)}} - 1} \right) + {\rm{ReLU}}\left( {R_i^{\tilde I,\tilde J}} \right)\left( {{{\tilde J}^{\left( i \right)}} - 1} \right)} \right)},
\end{aligned}
\end{equation}
\end{linenomath*}
and a representation similar to \eqref{4-15-2} for incompressible materials.
\begin{remark}
It is noted that the above normality terms for both compressible and incompressible materials have been derived under the assumption that all isotropic invariants, i.e., ${{{\rm{I}}_{\bf{C}}}},\, {{\rm{I}}{{\rm{I}}_{\bf{C}}}},\, {\rm{and}}\,\, {\rm{II}}{{\rm{I}}_{\bf{C}}}$, are available in the formulation. The case where only a subset of these invariants is considered can be addressed straightforwardly by a slight modification of the aforementioned terms.
\end{remark}
% \subsection{Growth Condition}
% In data-driven methods, enforcing the growth condition—either by constraining the structure of MLPs or multiplying predefined factors to those functions—can be challenging, especially when accounting for other requirements, including objectivity, material symmetry, polyconvexity, and normality. Alternatively, one can add analytical terms that satisfy such a condition, as suggested in 
%  \cite{klein2022polyconvex, linden2023neural}. In this regard, we use the following term \cite{hartmann2003polyconvexity}:
%  \begin{linenomath*}
% \begin{equation}\label{DD2}
% \begin{aligned}
% {{\tilde \psi }^{growth}}\left( {{\rm{II}}{{\rm{I}}_{\bf{C}}}} \right) = {(\sqrt {{\rm{II}}{{\rm{I}}_{\bf{C}}}}  + \frac{1}{{\sqrt {{\rm{II}}{{\rm{I}}_{\bf{C}}}} }} - 2)^2}, 
% \end{aligned}
% \end{equation}
% \end{linenomath*}
% which results in the following stress:
%  \begin{linenomath*}
% \begin{equation}\label{DD3}
% \begin{aligned}
% {{\bf{S}}^{growth}} = 2\sqrt {{\rm{II}}{{\rm{I}}_{\bf{C}}}} \left( {\sqrt {{\rm{II}}{{\rm{I}}_{\bf{C}}}}  + \frac{1}{{\sqrt {{\rm{II}}{{\rm{I}}_{\bf{C}}}} }} - 2} \right)\left( {1 - \frac{1}{{{\rm{II}}{{\rm{I}}_{\bf{C}}}}}} \right){{\bf{C}}^{ - 1}}.
% \end{aligned}
% \end{equation}
% \end{linenomath*}
\section{Training}
\begin{figure}[h!]
    \centering
    \makebox[\textwidth][c]{%
        \includegraphics[width=1.0\textwidth]{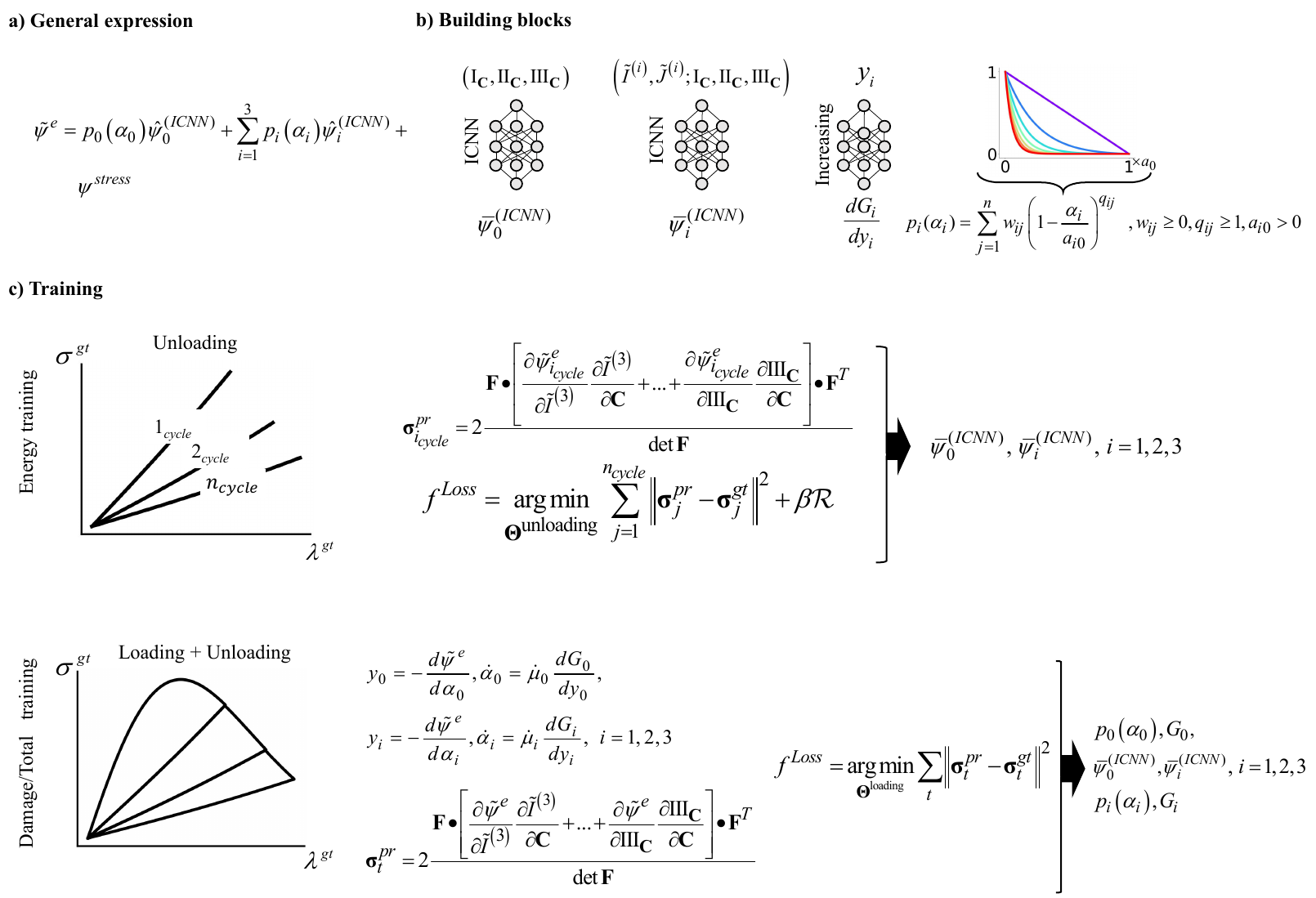}
    }
    \caption{(a) Generic architecture showing the strain-energy and attenuation functions $p_i(\alpha_i)$. (b) Constitutive building blocks include ICNN architectures for strain energy potentials, increasing dissipation through positive dissipation rates $G'_i(y_i)$, and convex decreasing degradation function expansion. (c) Two-stage learning protocol: Stage I fits unloading branches to calibrate the elastic energy, Stage II freezes the energy and optimizes damage evolution on full cyclic data. In either stage, the loss function is based on the comparison between the predicted and ground truth stresses.}
    \label{main_algo}
\end{figure}

To train the proposed data-driven model, we assume that stress-strain data are available. The parameters of the scalar potentials ${\bar \psi^{(ICNN)} _0}$, ${\bar \psi _i^{(ICNN)}}$, and $G_i$ for $i = 1, 2, 3$, along with the attenuation functions $p_i$, are optimized to minimize the discrepancy between the predicted stress vectors and the ground truth data. Also, although the above formulation retain the normalization condition for stresses and energies, but it is numerically beneficial to keep normalization coefficient defined above as small as possible. Hence, we define the following terms for compressible and in compressible materials, respectively:
 \begin{linenomath*}
\begin{equation}\label{T00}
\begin{aligned}
&{\cal R} = {\left( {R_0^{{{\rm{I}}_{\bf{C}}},{\rm{I}}{{\rm{I}}_{\bf{C}}},{\rm{II}}{{\rm{I}}_{\bf{C}}}}} \right)^2} + \sum\limits_{i = 1}^3 {\left[ {{{\left( {R_i^{\tilde I,\tilde J}} \right)}^2} + {{\left( {R_i^{{{\rm{I}}_{\bf{C}}},{\rm{I}}{{\rm{I}}_{\bf{C}}},{\rm{II}}{{\rm{I}}_{\bf{C}}},\tilde J}} \right)}^2}} \right]},\\
&{\cal R} = {\left( {R_0^{{{\rm{I}}_{\bf{C}}},{\rm{I}}{{\rm{I}}_{\bf{C}}}}} \right)^2} + \sum\limits_{i = 1}^3 {\left[ {{{\left( {R_i^{\tilde I,\tilde J}} \right)}^2} + {{\left( {R_i^{{{\rm{I}}_{\bf{C}}},{\rm{I}}{{\rm{I}}_{\bf{C}}}}} \right)}^2}} \right]}.
\end{aligned}
\end{equation}
\end{linenomath*}
These quantities are incorporated into the loss function during training. To formulate the training problem, let each pair of strain and stress data be denoted by $\left( \mathbf{F}_t^{\mathrm{gt}}, \boldsymbol{\sigma}_t^{\mathrm{gt}} \right)$, and the corresponding model predictions for the given ground truth strain be represented as $\left( \mathbf{F}_t^{\mathrm{gt}}, \boldsymbol{\sigma}_t^{\mathrm{pr}} \right)$. The objective is then to solve the following minimization problem:
 \begin{linenomath*}
\begin{equation}\label{T0}
\begin{aligned}
Loss = \mathop {\arg \,\min }\limits_{\bm{\Theta }} \sum\limits_t {{{\left\| {{\bf{\sigma }}_t^{pr} - {\bf{\sigma }}_t^{gt}}\right\|}^2}}+ \beta \mathcal{R}
\end{aligned}
\end{equation}
\end{linenomath*}
where $\beta$ is a hyper parameter, and  ${\bm{\Theta }}$ denotes the set of all parameters used in the data-driven method. Unlike non-dissipative models, the present data-driven framework is path-dependent, which introduces an explicit dependence on (pseudo) time. Another distinguishing feature is the presence of additional equations, governing evolution of the damage variables. These differences considerably increase the computational cost when all parameters are trained all at once. Therefore, it is desirable to adopt a training strategy that is more computationally efficient for the general case. In this regard, one can exploit the physical fact that occurs in the irreversible process. That is, during unloading, the state of the damage remains the same. Hence, one can use the unloading part of the data to achieve a good training for the strain energy densities where it is required to define attenuation factors to account for constant stage of the damage in each unloading cycles.  Hence, the following loss function for the training of the strain energy densities is considered in this study: 
 \begin{linenomath*}
\begin{equation}\label{T1}
\begin{aligned}
Loss = \mathop {\arg \,\min }\limits_{{{\bm{\Theta }}^{unloading}}} \sum\limits_{{t_u}} {{{\left\| {{\bf{\sigma }}_{{t_u}}^{pr} - {\bf{\sigma }}_{{t_u}}^{gt}} \right\|}^2}}+\beta \mathcal{R},  
\end{aligned}
\end{equation}
\end{linenomath*}
with 
\begin{linenomath*}
\begin{equation}\label{T11}
\begin{aligned}
{{\bm{\Theta }}^{unloading}} = \left\{{{\bm{\Theta }} _{{{\bar \psi }_{0}^{ICNN}}}},\,{{{\bm{\Theta }} _{\bar \psi _i^{ICNN}}},\,{{\mathcal P}_0}, {{\mathcal P}_i}} \right\},\,\,\, i=1,2,3.  
\end{aligned}
\end{equation}
\end{linenomath*}
in which ${{\cal P}_i} = \left[ {{p_i}\left( {{\alpha _{i,1}}} \right),\,...,\,{p_i}\left( {{\alpha _{i,{n_{cycle}}}}} \right)} \right]$ with  $n_{cycle}$ is the number of unloading existing in the stress-strain data. The only condition is:
\begin{linenomath*}
\begin{equation}\label{T2}
\begin{aligned}
1 > {p_i}\left( {{\alpha _{i,1}}} \right) > ... > {p_i}\left( {{\alpha _{i,{n_{cycle}}}}} \right) \ge 0.
\end{aligned}
\end{equation}
\end{linenomath*}
Similar expression and the corresponding condition \eqref{T2} hold for ${{\mathcal P}_0}$. Since the focus of minimization \eqref{T1} is to train strain energy densities, we follow a staggered strategy in this stage. In particular, we assume the coefficients used for normalization of the stress, i.e., $R_0^{{{\rm{I}}_{\bf{C}}},{\rm{I}}{{\rm{I}}_{\bf{C}}}},\,\,R_i^{{{\rm{I}}_{\bf{C}}},{\rm{I}}{{\rm{I}}_{\bf{C}}}},\,\,R_i^{\tilde I,\tilde J}$ for compressible and $R_0^{{{\rm{I}}_{\bf{C}}},{\rm{I}}{{\rm{I}}_{\bf{C}}}},\,\,R_i^{{{\rm{I}}_{\bf{C}}},{\rm{I}}{{\rm{I}}_{\bf{C}}}},\,\,R_i^{\tilde I,\tilde J}$ for incompressible materials, as non-trainable constants within the first term of \eqref{T1}, while updating them after each training epoch to ensure satisfaction of the normalization condition throughout the training process.  \\
After tuning parameters ${{\bm{\Theta }}^{unloading}}$, in the second training, we train the whole data for the following:
\begin{linenomath*}
\begin{equation}\label{T3}
\begin{aligned}
&{f^{Loss}} = \mathop {\arg \,\min }\limits_{{{\bm{\Theta }}^{loading}}} \sum\limits_t {{{\left\| {{\bf{\sigma }}_t^{pr} - {\bf{\sigma }}_t^{gt}} \right\|}^2}},\\
&{{\bm{\Theta }}^{loading}} = \left\{ {{{\bm{\Theta }} _{{p_0}}},\, {{\bm{\Theta }} _{{p_i}}},\, {{\bm{\Theta }} _{{G_0}}}, \,{{\bm{\Theta }} _{{G_i}}},\,{{\cal C}_0},\, {{\cal C}_i}} \right\},\,\, i=1,2,3.
\end{aligned}
\end{equation}
\end{linenomath*}
in which ${{{\bm{\Theta }} _{{p_i}}}}$, ${{{\bm{\Theta }} _{{G_i}}}}$, and ${\mathcal{C}}_i$ $i=1,2,3$ are, respectively, parameters of $p_i$, $G_i$, and constants of energy that are defined as follows:
\begin{linenomath*}
\begin{equation}\label{T4}
\begin{aligned}
&{\tilde \psi ^e} = \frac{{\tilde \psi _0^{(ICNN)}}}{{{{\cal C}_0}}} + \sum\limits_{i = 1}^3 {\frac{{\tilde \psi _i^{(ICNN)}}}{{{{\cal C}_i}}}} ,\\
&{\mathcal {C}}_0>0,\, {\mathcal {C}}_i>0, i=1,2,3.
\end{aligned}
\end{equation}
\end{linenomath*}
Indeed, we define ${\mathcal{C}}$'s in the second part since, for example for $i=1,2,3$,  in the first training part \eqref{T1} for any given set of trained parameters, the combination
\begin{linenomath*}
\begin{equation}\label{T5}
\begin{aligned}
&\gamma_i{{\cal P}_i} = \left[ \gamma_i{{p_i}\left( {{\alpha _{i,1}}} \right),\,...,\,\gamma_i{p_i}\left( {{\alpha _{i,{n_{cycle}}}}} \right)} \right]\\
& {\frac{1}{\gamma_i} {\tilde \psi _i} }, \gamma_i \in {\mathbb{R}}^+, i=1,2,3,
\end{aligned}
\end{equation}
\end{linenomath*}
for every $\gamma_i$ can be considered a solution for the optimization \eqref{T0}. Hence, in the second training stage, we consider \eqref{T4} and train the scaling factors along with the parameters of the damage potential and attenuation functions. Apparently, the same argument is valid for ${\tilde \psi _0^{(ICNN)}}$. The main steps of the new algorithm have been illustrated in Fig. \ref{main_algo}.   
It is worth noting that if the desired accuracy is not achieved with the decoupled algorithm, a brief round of full-parameter training, as formulated in \ref{T0}, is necessary.  
\section{Numerical examples}
In this section, we assess the capability of the proposed framework by considering several examples. The training data is only the stress-strain curves for all synthetic and the experimental cases.\\ In the first example, we test the method for capturing the degradation in an isotropic material due to isotropic damage. For all synthetic materials, the following  closed-form  damage potentials are used 
\cite{noel2019modeling,ostwald2019implementation,li2016damage} (see Fig. \ref{appen}):
\begin{linenomath*}
\begin{equation}\label{N1}
\begin{aligned}
&{G_{{\text{Noel}}}} = \exp \left( { - \exp \left( {10\left( {0.2 - r} \right)} \right)} \right),\\
&{G_{{\text{Oswalt}}}} = 1.0 - {\text{ }}\exp \left( { - 4\left( r \right)} \right),\\
& {G_{Li}} = (1.0 - {\text{ exp}}\left( { - \frac{r}{2}} \right)).
\end{aligned}
\end{equation}
\end{linenomath*}
The energy density employed for the generation of the synthetic data is:
\begin{linenomath*}
\begin{equation}\label{N2}
\begin{aligned}
\psi^{gt}  = \left( {1 - \alpha } \right)\frac{\mu }{2}({{\rm{I}}_{\bf{C}}} - 3) + {P_{pre}}\left( {\sqrt {{\rm{II}}{{\rm{I}}_{\bf{C}}}}  - 1} \right).
\end{aligned}
\end{equation}
\end{linenomath*}
We test the isotropic version of the data-driven model:
\begin{linenomath*}
\begin{equation}\label{N3}
\begin{aligned}
&{\psi ^{pr}} = \tilde \psi _0^{(ICNN)}({{\rm{I}}_{\bf{C}}},\,{\rm{I}}{{\rm{I}}_{\bf{C}}},{\alpha _0}) + {P_{pre}}\left( {\sqrt {{\rm{II}}{{\rm{I}}_{\bf{C}}}}  - 1} \right),\\
&\dot \alpha_0  = \dot y\frac{{dG_0}}{{dy_0}},\,\,\, \dot y \ge 0.
\end{aligned}
\end{equation}
\end{linenomath*}

\begin{figure}[h!]
    \centering
        \begin{adjustbox}{center}
        \includegraphics[width=0.9\textwidth]{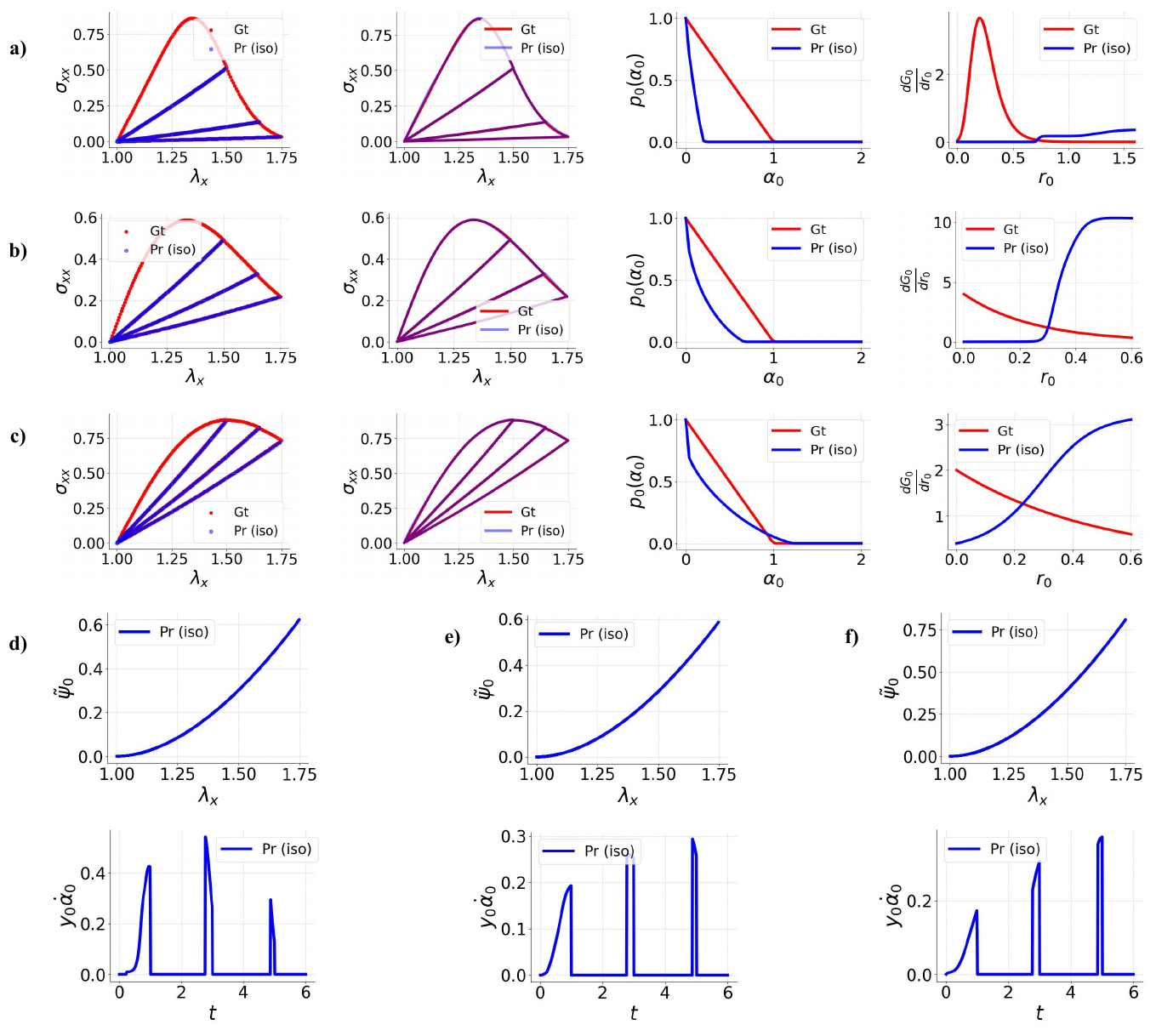}
    \end{adjustbox}
    \caption{Validation of the model in recovering synthesized isotropic materials. The synthesized materials were generated based on models $\eqref{N1}_1$, $\eqref{N1}_2$, and $\eqref{N1}_3$, with comparisons to the trained data-driven models shown in panels (a), (b), and (c), respectively. The decoupled training scheme described in algorithm~\ref{main_algo} was employed. Each panel presents results from the energy and damage training stages and includes comparisons of the attenuation functions and damage evolution functions between the synthesized materials and the corresponding trained models. The non-negativity of the resulting strain energy density functions and positivity of dissipation rates for all data-driven models were plotted in (d), (e), and (f), corresponding respectively to synthesized materials with damage evolution functions $\eqref{N1}_1$, $\eqref{N1}_2$, and $\eqref{N1}_3$.}
    \label{iso_model_iso_damage} 
\end{figure}

 The architectures used in this example are $[2,\,3,\,3,\,1]$ and $[1,\,7,\,7,\,1]$ to approximate $\bar\psi^{(ICNN)}$ and $G'^{\,\,\, (MLP)}$, respectively. The values $[1,\, 1.5,\, 2.0,\, 3.0,\,5.0,\, 7.0,\,10.0,\,15.0,\, 20.0,\, 30.0,\,50.0,\,70.0,\,100.0,\,150,\,200.0]$ correspond to the $q_j$'s defined in  \eqref{DD4}.   Figs. $\ref{iso_model_iso_damage}$a \& $\ref{iso_model_iso_damage}$d, $\ref{iso_model_iso_damage}$b \& $\ref{iso_model_iso_damage}$e, and $\ref{iso_model_iso_damage}$c \& $\ref{iso_model_iso_damage}$f show the results when $G_{\textnormal{Noel}}$, $G_{\textnormal{Oswalt}}$, and $G_{\textnormal{Li}}$ are used for generation of the ground truth data, respectively. As can be seen, for training the data, we used the decoupled scheme, recovering first the unloading stress response by training the strain energy density $\bar\psi^{(ICNN)}$ along with some constants attenuation values, then the entire loading history by training the damage potential and the attenuation functions along with constants used to scale the energy. The method is robust regardless of the damage model, showing the effectiveness of the proposed training strategy and the expressiveness of data-driven setup. From these figures, the model recovered the stress-strain curve extremely accurate. Note that we use degradation functions $p_0\left(\alpha_0\right)$ that are more expressive than the traditional $[1-\alpha]$ approach. In fact, the fitted degradation functions differ from the generating model $[1-\alpha]$. The same conclusion is true for $G$ functions. These results clearly show the non-unique nature of the problem, confirming the existence of multiple states of internal variables and their evolution for a given stress–strain dataset. Furthermore, according to the numerical results shown in Figs $\ref{iso_model_iso_damage}$d, $\ref{iso_model_iso_damage}$e, and $\ref{iso_model_iso_damage}$f, the trained strain potential energies are non-negative in the training range.  Also,  as we obtain the stress as the derivative of the Helmholtz potential, we numerically showed the satisfaction of the dissipation inequality in Figs $\ref{iso_model_iso_damage}$d, $\ref{iso_model_iso_damage}$e, and $\ref{iso_model_iso_damage}$f for all synthetic models to indicate that the new data-driven method is thermodynamically consistent according to the definition \ref{th_consistency}.

\begin{figure}[h!]
    \centering
    \begin{adjustbox}{center}
        \includegraphics[width=0.9\textwidth]{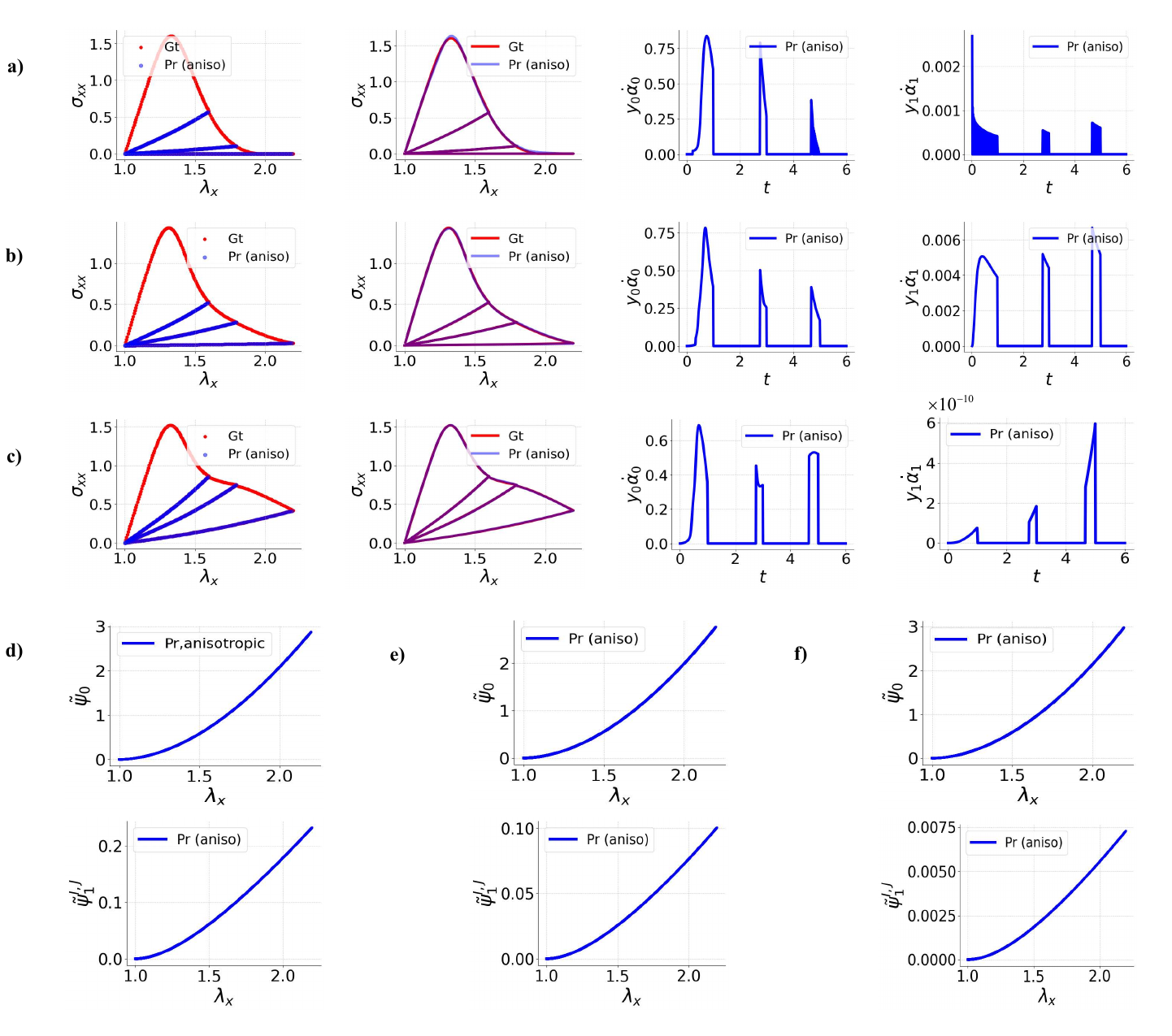}
    \end{adjustbox}
    \caption{Validation of the model in recovering the damage behavior of synthesized transversely isotropic materials. The synthesized materials were generated based on models $\eqref{N1}_1$, $\eqref{N1}_2$, and $\eqref{N1}_3$ for anisotropic part and $\eqref{N1}_1$ for the isotropic one. The decoupled training scheme described in algorithm~\ref{main_algo} was employed. Panels (a), (b), and (c) show the results of the energy and damage training stages, as well as the dissipation rates for the three data-driven models. The non-negativity of the resulting strain energy density functions is shown in panels (d), (e), and (f), corresponding respectively to the synthesized materials with damage evolution functions $\eqref{N1}_1$, $\eqref{N1}_2$, and $\eqref{N1}_3$ for anisotropic part.}
    \label{transversely_iso_damage} 
\end{figure}
% In the training, as we have three unknown functions, we performed the training for all functions at the same time. As can be seen, the proposed model captured the stress prediction for both model damages. However, for the case of damage growth, we can see that the model obtained a path different from the original one for the second damage model, although the corresponding stresses are almost the same. However, this is not a problem as we trained the model based on the stress-strain curve without knowledge about the damage rule. This result indicates that there is no unique constitutive relation for a given stress-strain data with damage. \\

\begin{figure}[h!]
    \centering
    \begin{adjustbox}{center}
        \includegraphics[width=0.9\textwidth]{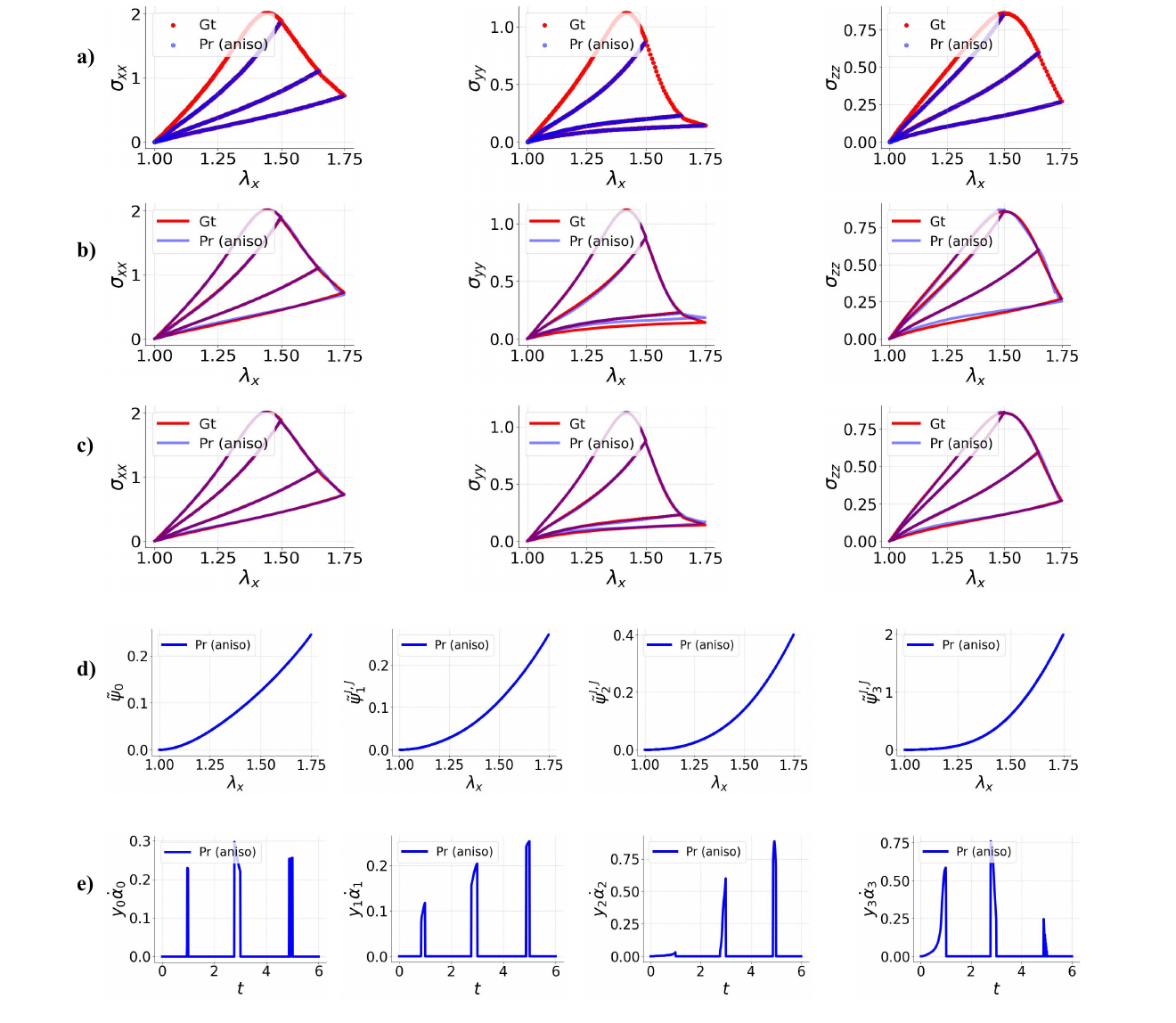}
    \end{adjustbox}
    \caption{Evaluation of the capability of the proposed data-driven model to predict the damage behavior of a synthesized orthotropic compressible material under a uniaxial test. The training was performed using all three resulting axial stress components. Panels (a) and (b) show the results of the energy and damage training stages, respectively, based on the proposed decoupled training procedure. Panel (c) shows the results of training when all parameters are trained together. Panel (d)  demonstrates the non-negativity of the defined strain energy density functions, while panel (e) shows the positivity of the corresponding dissipation rates. }
    \label{comp_damage} 
\end{figure}

In the next example, we consider the anisotropy by considering a transversely isotropic material. For the synthesized data, we use the following synthetic model:
\begin{linenomath*}
\begin{equation}\label{N4}
\begin{aligned}
 \psi^{gt} =& \left( {1 - {\alpha _0}} \right)\left[ {\frac{1}{2}\left( {{{\left( {\frac{{{{\rm{I}}_{\bf{C}}}}}{3}} \right)}^2} - 1} \right) + \frac{1}{3}\left( {{{\left( {\frac{{{\rm{I}}{{\rm{I}}_{\bf{C}}}}}{3}} \right)}^3} - 1} \right)} \right]+\\
& \left( {1 - {\alpha _1}} \right)\left[ {\frac{1}{3}\left( {{{\left( {{{\tilde I}^{\left( 1 \right)}}} \right)}^{1.5}} - 1} \right) + \frac{1}{5}\left( {{{\left( {{{\tilde J}^{\left( 1 \right)}}} \right)}^{2.5}} - 1} \right)} \right] + {P_{pre}}\left( {\sqrt {{\rm{II}}{{\rm{I}}_{\bf{C}}}}  - 1} \right),\\
&{{{\mathbf{\tilde L}}}_1} = \left[ {\begin{array}{*{20}{c}}
  {0}&0&0 \\ 
  0&{0.5}&0 \\ 
  0&0&{0.5} 
\end{array}} \right].
\end{aligned}
\end{equation}
\end{linenomath*}
For the evolution of $\alpha_0$,  we use $\eqref{N1}_1$in all examples, while $\eqref{N1}_1$, $\eqref{N1}_2$, and $\eqref{N1}_3$ are utilized for the evolution of $\alpha_1$ to generate ground truth data in Fig. $\ref{transversely_iso_damage}$a,  Fig. $\ref{transversely_iso_damage}$b, and Fig. $\ref{transversely_iso_damage}$c, respectively. For the data-driven model, we consider the following form:
\begin{linenomath*}
\begin{equation}\label{N5}
\begin{aligned}
{\psi ^{pr}} = \tilde \psi _0^{(ICNN)}({{\rm{I}}_{\bf{C}}},{\rm{I}}{{\rm{I}}_{\bf{C}}},{\alpha _0}) + \tilde \psi _1^{(ICNN)}({\tilde I^{\left( 1 \right)}},{\tilde J^{\left( 1 \right)}},{\alpha _1}) + {P_{pre}}\left( {\sqrt {{\rm{II}}{{\rm{I}}_{\bf{C}}}}  - 1} \right).
\end{aligned}
\end{equation}
\end{linenomath*}
The architectures used in this example are $[2,\,3,\,3,\,1]$ and $[1,\,7,\,7,\,1]$ to approximate $\bar\psi^{(ICNN)}$'s and $G'^{\,\,\, (MLP)}$'s, respectively. Additionally, the values $[1.0,\, 1.5,\, 1.75,\, 2.0,\, 4.0,\,6.0,\,8.0,\, 10.0,\, 15.0,\, 20.0,\, 40.0,\,80.0,\,100.0,\,150.0,\, 200.0]$ are considered for the $q_j$'s defined in \eqref{DD4} in the data-driven models used to predict ground truth data shown in Figs. \ref{transversely_iso_damage}a \& \ref{transversely_iso_damage}b, while the values $[1,\, 1.5, \,1.75,\, 2.0,\, 2.5,\,6.0,\,7.5,\, 10.0,\, 15.0,\, 20.0,\, 40.0,\,80.0,\,100.0,\,150.0,\, 200.0]$ in the model corresponding to Fig.\ref{transversely_iso_damage}c. We show the results in Figs. $\ref{transversely_iso_damage}$a \& $\ref{transversely_iso_damage}$d, $\ref{transversely_iso_damage}$b \& $\ref{transversely_iso_damage}$e, and $\ref{transversely_iso_damage}$c \& $\ref{transversely_iso_damage}$f. Similarly to the isotropic example, we used the decoupled training scheme elaborated earlier. As can be seen, there is perfect match between the predicted data and train data for both the unloading stresses and the full loading cycles. As discussed earlier, for the decoupled scheme, the unloading portion of the data informs the strain energy densities, while the degradation function and damage evolution equations are fitted in the second stage, when all the loading cycle is considered. Similar to the isotropic example, one can observe from Figs. \ref{transversely_iso_damage}d, \ref{transversely_iso_damage}e, and \ref{transversely_iso_damage}f that the trained strain energy potentials are non-negative, and the method remains thermodynamically consistent in the sense of Definition \ref{th_consistency}.Moreover, it is noted from Figs.\ref{transversely_iso_damage}c and \ref{transversely_iso_damage}f that, while the ground-truth model exhibits anisotropic damage, the predicted damage model mostly relies on the isotropic contribution. This can be attributed to  the nonlinear format of the attenuation functions, combined with the nonlinear strain energy densities and damage potentials, results in a more expressive model that can approximate the anisotropic behavior using a predicted isotropic one when there is no multi-axial data available.\\
As the next example, we test the new data-driven damage method to recover the stress-strain curve of a synthetic compressible material. To generate the ground truth stress-strain data, we consider the following Helmholtz potential: 

\begin{linenomath*}
\begin{equation}\label{N6-0}
\begin{aligned}
& \psi^{gt} = \frac{{1 - {\alpha _0}}}{5}\left( {\frac{{{{\rm{I}}_{\bf{C}}}{\rm{ + I}}{{\rm{I}}_{\bf{C}}}}}{3}{\rm{ + III}}_{\bf{C}}^{ - 1} - 3} \right) + \frac{1}{{20}}\sum\limits_{i = 1}^3 {\left( {1 - {\alpha _i}} \right)\left[ {\frac{{{{\left( {{{\tilde I}^{\left( i \right)}}} \right)}^{i + 1}} - 1}}{{i + 1}} + \frac{{{{\left( {{{\tilde J}^{\left( i \right)}}} \right)}^{i + 1}} - 1}}{{i + 1}}{\rm{ + III}}_{\bf{C}}^{ - 1} - 1} \right]}, \\
&{{{\mathbf{\tilde L}}}_1} = \left[ {\begin{array}{*{20}{c}}
  {1.0}&0&0 \\ 
  0&0&0 \\ 
  0&0&0 
\end{array}} \right],\,\,{{{\mathbf{\tilde L}}}_2} = \left[ {\begin{array}{*{20}{c}}
  0&0&0 \\ 
  0&{1.0}&0 \\ 
  0&0&0 
\end{array}} \right],\,\,{{{\mathbf{\tilde L}}}_3} = \left[ {\begin{array}{*{20}{c}}
  0&0&0 \\ 
  0&0&0 \\ 
  0&0&{1.0} 
\end{array}} \right],
\end{aligned}
\end{equation}
\end{linenomath*}
where $\eqref{N1}_1$ was used for the damage evolutions. To obtain the data, we considered a uniaxial test that results in all three axial stresses. In this example, we use the following format as the data-driven form:
\begin{linenomath*}
\begin{equation}\label{N5-1}
\begin{aligned}
&{\psi ^{pr}} = \tilde \psi _0^{(ICNN)}({{\rm{I}}_{\bf{C}}},{\rm{I}}{{\rm{I}}_{\bf{C}}},{\rm{II}}{{\rm{I}}_{\bf{C}}},{\alpha _0}) + \sum\limits_{i = 1}^3 {\tilde \psi _i^{(ICNN)}({{\tilde I}^{\left( i \right)}},{{\tilde J}^{\left( i \right)}},{\rm{II}}{{\rm{I}}_{\bf{C}}},{\alpha _i})},
\end{aligned}
\end{equation}
\end{linenomath*}
along with \eqref{DD5}. The architectures employed in this example are $[3,\,3,\,3,\,1]$ and $[1,\,7,\,7,\,1]$ to approximate $\bar \psi^{(ICNN)}$'s and $G'^{\,\,(MLP)}$'s, respectively. The same values as in the isotropic example are used for the $q_j$'s defined in equation \eqref{DD4}. We trained the model against the data for all three axial stresses. The results have been illustrated in Fig. \ref{comp_damage}. Similar to previous examples, we applied the decoupled training scheme. In Fig. \ref{comp_damage}a, the results of the training for the unloading part for strain energy potential for all three stresses, showing a perfect match between the trained model and the synthetic ground truth data. The results corresponding to the second stage of the decoupled algorithm are shown in Fig.  \ref{comp_damage}b. As can be seen, minor discrepancies remain after the decoupled training for all data, which are fully resolved by the final training of all parameters, as illustrated in Fig.\ref{comp_damage}c. During the final stage of training all parameters, the best-performing result was selected.   Moreover, as illustrated in Fig.  \ref{comp_damage}d, the trained strain energies are non-negative, analogous to previous synthetic examples.  Furthermore, Fig. \ref{comp_damage}e confirms that the new data-driven method is thermodynamically consistent.\\

\begin{figure}[h!]
    \centering
    \begin{adjustbox}{center}
        \includegraphics[width=0.9\textwidth]{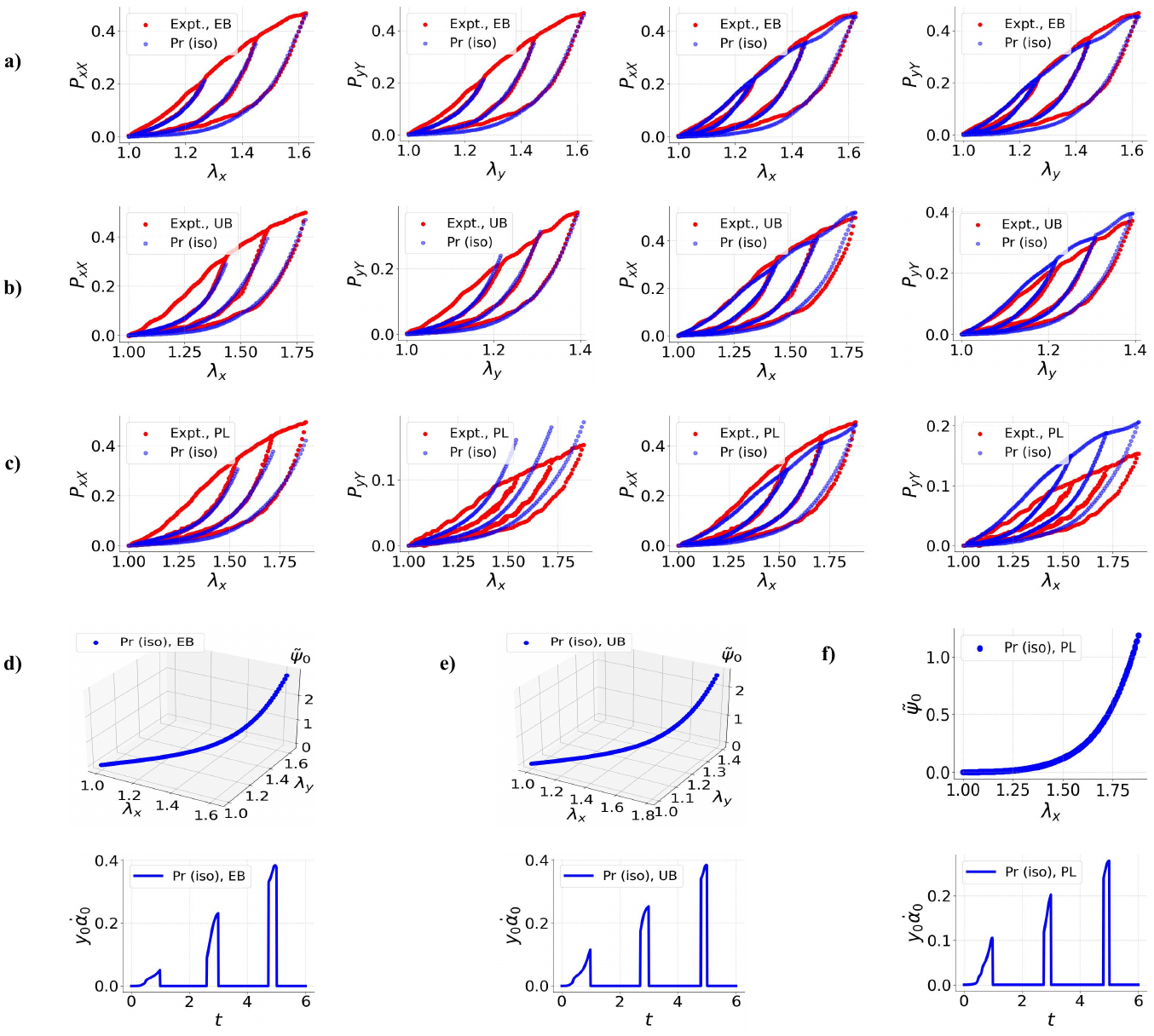}
    \end{adjustbox}
    \caption{The isotropic data-driven damage model was used to predict the nominal stress for various experimental datasets corresponding to double-network hydrogels, as reported in \cite{mai2018distinctive}. The material is assumed to be incompressible and initially isotropic. Panels (a), (b), and (c) compare the model predictions with experimental results under equi-biaxial, unequal biaxial, and planar loading conditions, respectively. The isotropic model partially captured the experimental responses, with greater discrepancies observed in the planar loading case, particularly in the stress components along the y-direction compared to the x-direction. Additionally, the trained strain energy density functions and dissipation rates associated with the isotropic data-driven model are shown in panels (d), (e), and (f), corresponding to the equi-biaxial, unequal biaxial, and planar loading conditions, respectively.}
    \label{expe_iso} 
\end{figure}

\begin{figure}[h!]
    \centering
    \begin{adjustbox}{center}
        \includegraphics[width=0.9\textwidth]{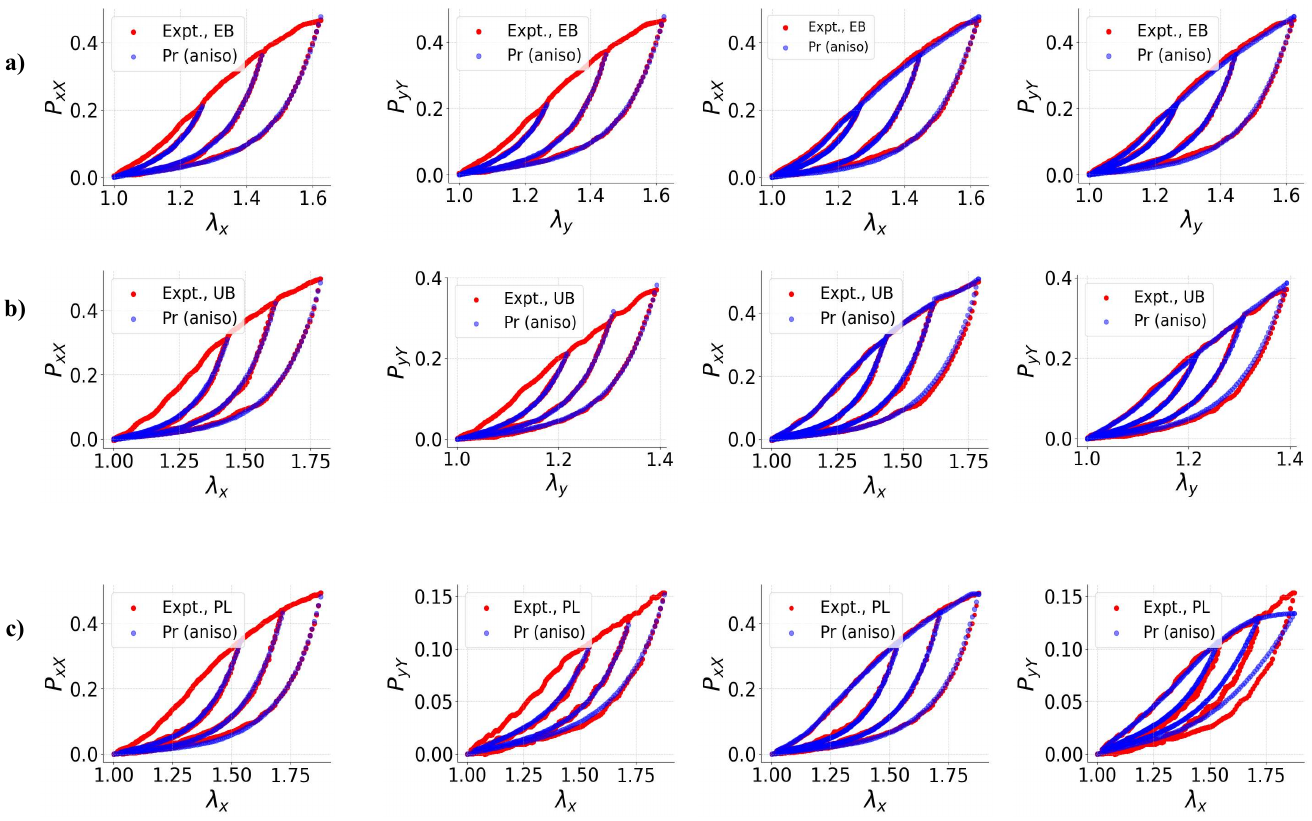}
    \end{adjustbox}
    \caption{The performance of the new anisotropic Data-driven damage model to predict the nominal stresses for various experimental datasets corresponding to double-network hydrogels, as reported in \cite{mai2018distinctive}. Panels (a), (b), and (c) compare the model predictions with experimental results under equi-biaxial, unequal biaxial, and planar loading conditions, respectively.The advantage of the anisotropic model in predicting the constitutive behavior of initially isotropic double-network materials is evident when compared to the isotropic damage model, highlighting the appearance of anisotropy induced by damage.  }
    \label{expe_aniso1} 
\end{figure}

\begin{figure}[h!]
    \centering
    \begin{adjustbox}{center}
        \includegraphics[width=0.9\textwidth]{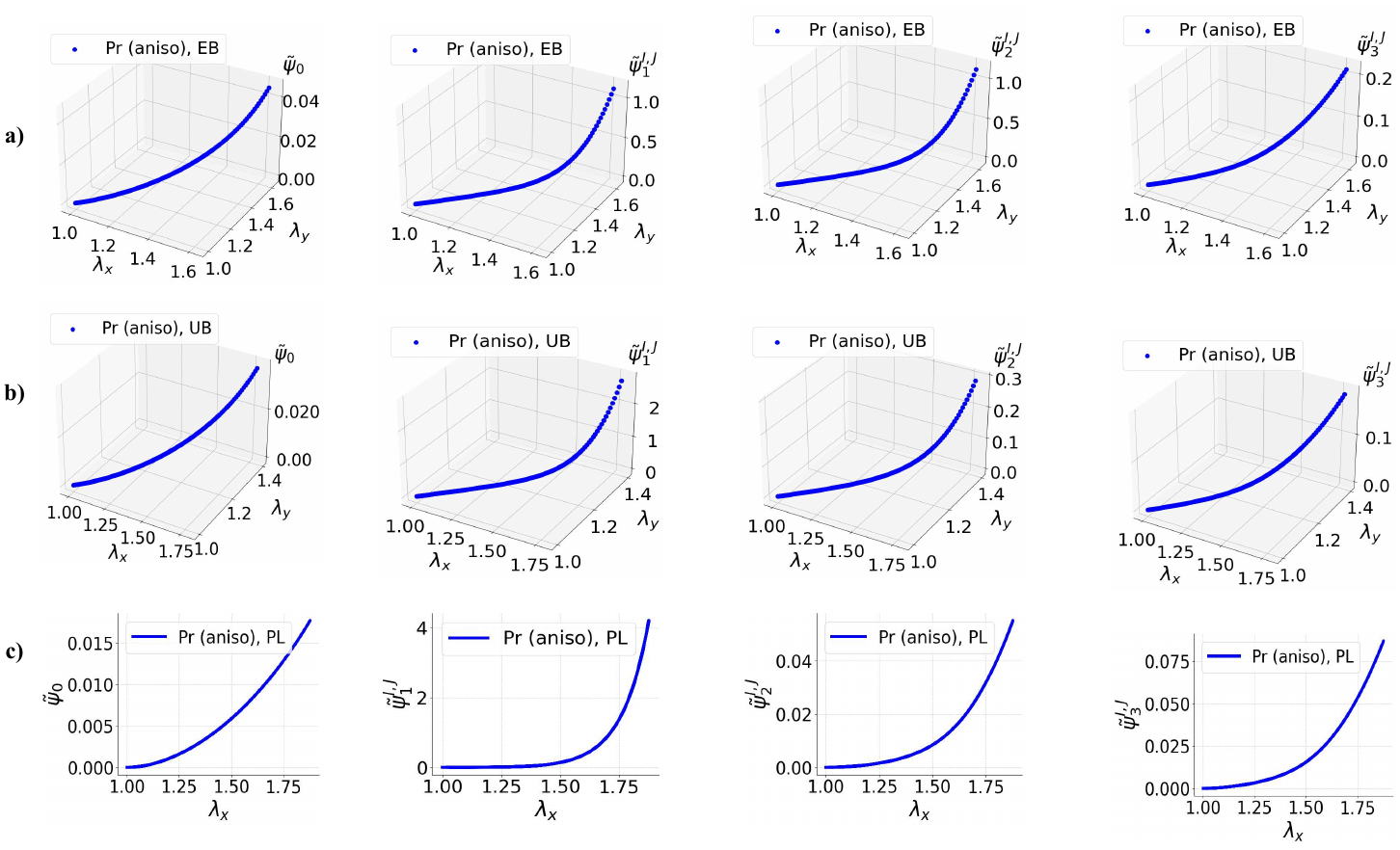}
    \end{adjustbox}
    \caption{The trained strain energy density functions corresponding to the anisotropic data-driven damage model \eqref{inc-3} were used to predict the experimental response of double-network hydrogels, as reported in \cite{mai2018distinctive}. Panels (a), (b), and (c) correspond to equi-biaxial, unequal biaxial, and planar loading conditions, respectively. }
    \label{expe_aniso2} 
\end{figure}

\begin{figure}[h!]
    \centering
    \begin{adjustbox}{center}
        \includegraphics[width=0.9\textwidth]{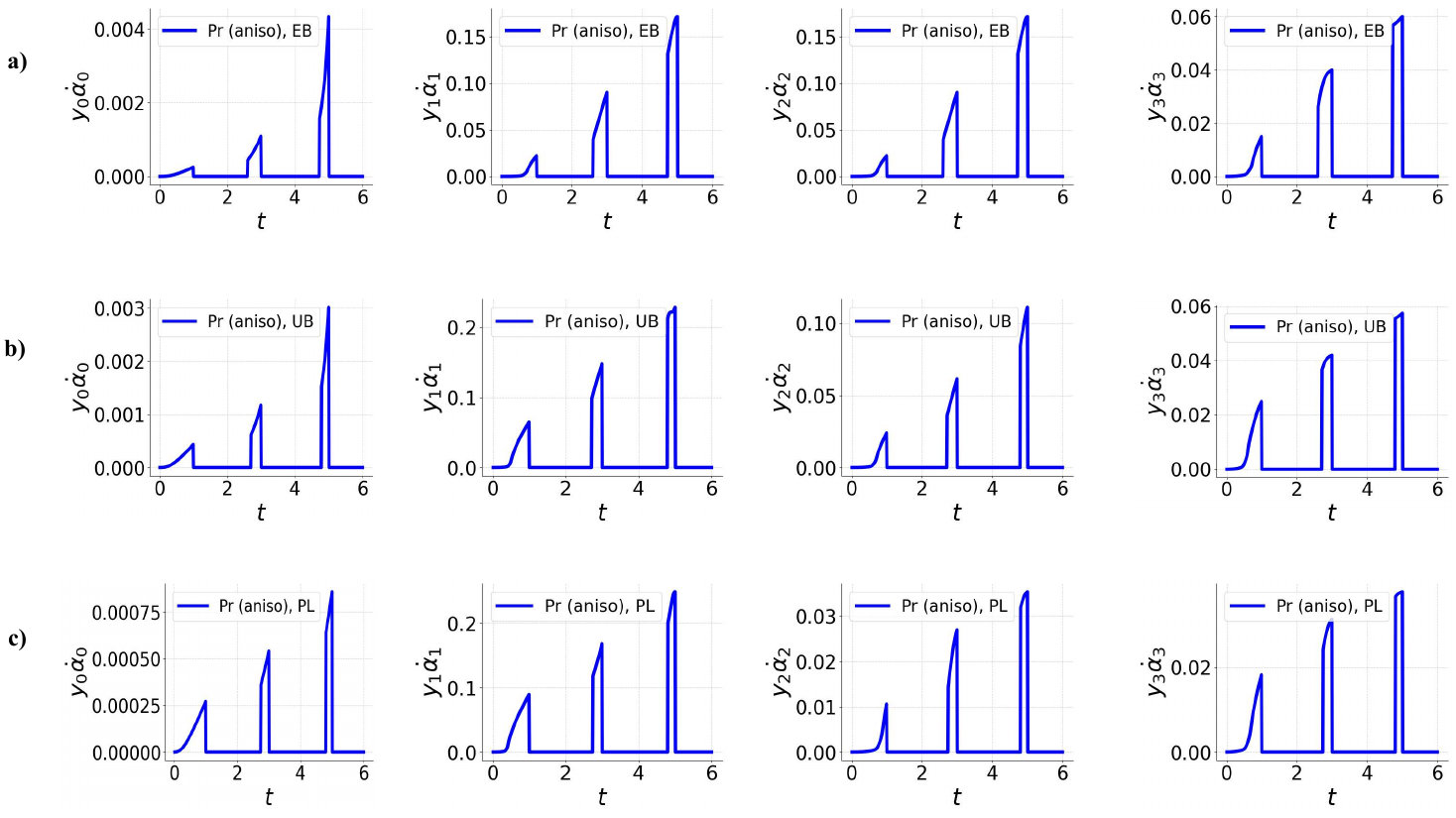}
    \end{adjustbox}
    \caption{The dissipative energy predicted by the trained data-driven model \eqref{inc-3} for double-network hydrogels, as reported in \cite{mai2018distinctive}, is illustrated. Panels (a), (b), and (c) correspond to equi-biaxial, unequal biaxial, and planar loading conditions, respectively. }
    \label{expe_aniso3} 
\end{figure}

As a last example, we consider the method in predicting experimental data reported in \cite{mai2018distinctive} for double network hydrogels. The data correspond to equi-biaxial, unequi-biaxial, and planar experiments. In particular, the state of strain can be identified as follows:
\begin{linenomath*}
\begin{equation}\label{N6}
\begin{aligned}
&{\rm{equal - biaxial}}:\,\,\,\,{\lambda _x} = {\lambda _y} = \lambda ,\,\,{\lambda _z} = \frac{1}{{{\lambda ^2}}},\\
&{\rm{unequal - biaxial}}:\,\,\,\,{\lambda _x} = \lambda ,\,\,{\lambda _y} = \frac{{1 + \lambda }}{2},\,{\lambda _z} = \frac{2}{{\lambda \left( {1 + \lambda } \right)}},\\
&{\rm{planar}}:\,\,\,\,{\lambda _x} = \lambda ,\,\,{\lambda _y} = 1.0,\,\,{\lambda _z} = \frac{1}{\lambda }.
\end{aligned}
\end{equation}
\end{linenomath*}
To highlight the effectiveness of anisotropic formulation, we consider two models. For the first model, we train an isotropic damage data-driven model, similar to \eqref{N3}, against all data. For the second model, on the other hand, we consider the general format suggested in \eqref{inc-3}, along with $\eqref{D15}_1$ and \eqref{D20}. The architectures of neural networks and $q_j$'s for isotropic and anisotropic models are chosen to be as similar as possible. In particular, we consider network of size $[2,3,3,1]$ for $\bar \psi ^{(ICNN)}_0$ and size of $[4,3,3,1]$ for anisotropic part $\bar \psi ^{(ICNN)}_1$. $[1,\,7,\,7,\,1]$ is employed to approximate all $G'^{\,\,(MLP)}$'s. The same values as in the isotropic example are used for the $q_j$'s defined in equation \eqref{DD4} for both isotropic and anisotropic models. To obtain the results, we first applied the proposed decoupled training procedure, then performed full-parameter training for both the isotropic and anisotropic models. To ensure a fair comparison, similar hyperparameters were used across models, and the parameters yielding the lowest loss corresponding to full-parameter training were selected.  The results for the isotropic and anisotropic cases are shown in Figs. \ref{expe_iso} and \ref{expe_aniso1}, respectively. Figs.\ref{expe_iso}a, \ref{expe_iso}b, and \ref{expe_iso}c present the results for the isotropic model under equal-biaxial, unequal-biaxial, and planar modes of deformation, respectively. Figs.\ref{expe_aniso1}a, \ref{expe_aniso1}b, and \ref{expe_aniso1}c show the corresponding results for the anisotropic model. The resulting strain energy densities and dissipation rates for the isotropic model are shown in Figs.\ref{expe_iso}d, \ref{expe_iso}e, and \ref{expe_iso}f, for equal-biaxial, unequal-biaxial, and planar modes, respectively. For the anisotropic model, the trained strain energy densities for the equal-biaxial, unequal-biaxial, and planar modes of deformation are shown in Figs.\ref{expe_aniso2}a, \ref{expe_aniso2}b, and \ref{expe_aniso2}c, respectively. The corresponding dissipation rates for these modes of deformations are illustrated in Figs.\ref{expe_aniso3}a, \ref{expe_aniso3}b, and \ref{expe_aniso3}c. For both isotropic and anisotropic models, the reported results in the unloading paths are those obtained from the first stage of the decoupled training, while the results for the loading–unloading paths correspond to the final stage, after training all parameters jointly.   For the isotropic model, it is observed that the model falls short to predict multi-axial results. In particular, although the results are acceptable for the case of equal-biaxial, the dependency is obvious for unequal-biaxial and planar cases, specifically for the stress in the $y$ direction. On the other hand, for anisotropic model,  one can see the model appropriately predicted the experimental results for all modes and all directions. These results clearly indicate that the anisotropic model outperforms isotropic models in predicting damage mechanism such as Mullin's effect.Also, from the results, it is clear that that both isotropic and anisotropic models are thermodynamically consistence in the sense of definition \ref{th_consistency}. 

\section{Concluding remarks}
In this study, we introduced a new data-driven anisotropic damage model for both compressible and incompressible orthotropic materials. The formulation is thermodynamically consistent, as designed theoretically and shown with the numerical results. The strain energy potentials are constructed to preserve polyconvexity—a condition which, together with an appropriate growth condition, ensures ellipticity. Additionally, the framework satisfies the normality condition through the inclusion of extra terms that respect polyconvexity.\\
Furthermore, numerical results across all examples demonstrate that the resulting strain energy density functions are non-negative. The proposed method is no longer constrained by the classical $[1-d]$ degradation function; instead, the attenuation of the Helmholtz energy is achieved by introducing a nonlinear function composed of a summation over a family of convex, decreasing functions. As a result, in contrast to traditional $[1-d]$ degradation, the new method offers more flexibility. This flexibility is further harnessed through the use of appropriately constrained neural networks for the dissipation potential and strain energy potentials.\\
For training, we proposed a decoupled scheme that significantly reduces the computational cost. Specifically, using either synthetic or experimental data, the strain energy density functions corresponding to the unloading branches of the stress–strain data are first trained. Subsequently, the parameters governing damage evolution are optimized using the entire dataset. If the mismatch between the predictions and ground truth data remains unsatisfactory, the full set of parameters can be jointly fine-tuned for a limited number of epochs. We tested this approach on multiple synthetic examples as well as one experimental dataset. The results demonstrate that this staged training strategy is both effective and scalable, especially in comparison to training all parameters simultaneously from the outset.\\
For the numerical examples, we considered three synthetic cases and one experimental dataset involving stress–strain curves for double-network hydrogels. The results confirm that the model is expressive enough to accurately capture a wide range of material behaviors. Additionally, we observed that the isotropic damage model fails to reproduce certain experimental data—particularly those from planar tests—whereas the anisotropic model successfully captures the dominant characteristics of the response.

In summary, we have presented a physics-augmented, data-driven framework to capture both inherent and damage-induced anisotropy in soft materials. By enforcing objectivity, thermodynamic consistency, and normality by construction, and by introducing a two-stage training strategy that isolates elastic and dissipative information, the model learns accurate, physically admissible constitutive behavior from limited multiaxial data. Synthetic benchmarks demonstrate robustness across isotropic, transversely-isotropic, and fully orthotropic responses, while application to tough double-network hydrogels highlights the practical gains of modeling damage-induced anisotropy. Future work will extend the approach to fracture, and embed the learned models in large-scale finite-element simulations of soft-tissue failure.
\section{Appendix}
\begin{figure}[h!]
    \centering
    \begin{adjustbox}{center}
        \includegraphics[width=.9\textwidth]{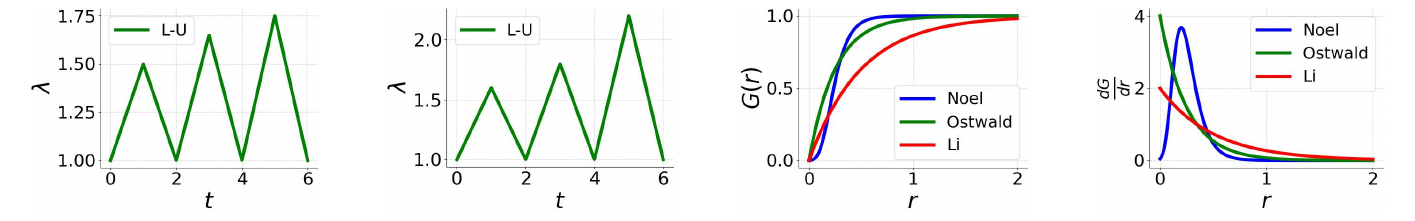}
    \end{adjustbox}
    \caption{The components used to generate the synthetic data. From left to right: the first plot shows the loading–unloading path employed to generate the ground-truth data in Figs.\ref{iso_model_iso_damage} and\ref{comp_damage}. The second plot shows the loading–unloading path used for generating the ground-truth data in Fig.\ref{transversely_iso_damage}. The last two plots show the analytical forms of $G$'s in \eqref{N1} and the corresponding derivatives  used in all synthetic examples.      }
    \label{appen} 
\end{figure}

The plots corresponding to the analytical damage potentials and their derivative alongside with the time-wise loading-unloading employed to compute ground truth data for synthetic examples are shown in Fig. \ref{appen}. 

\section*{Declarations}
The authors have no conflicts of interest to declare. 
\section*{Acknowledgments}
The authors gratefully acknowledge support from the U.S. Army Research Office under Award W911NF-24-1-0244.

\end{document}